\documentclass[aps,pre,reprint,groupedaddress,showpacs]{revtex4-1}
\usepackage{subfigure,epsfig,amsfonts,dcolumn}
\usepackage{amsmath}
\usepackage{amssymb}
\usepackage{amsthm}
\usepackage{multirow} 

\begin{document}

\title{Force propagation in isostatic granular packs}
\author{Nathan W. Krapf}
\affiliation{James Franck Institute, University of Chicago, 929 E. 57th Street, Chicago IL 60637}
\date{\today}

\newcommand{\nkv}[1]{\textbf #1}
\newcommand{\nkh}[1]{\hat{\nkv #1}}

\begin{abstract}
We investigate how forces spread through frictionless granular packs at the jamming transition. Previous work has indicated that such packs are isostatic, and thus obey a null stress law which, independent of the packing history, causes rays of stress to propagate away from a point force at oblique angles. Prior verifications of the null stress law have used a sequential packing method which yields packs with anisotropic packing histories. We create packs without this anisotropy, and then later break the symmetry by adding a boundary. Our isotropic packs are very sensitive, and their responses to point forces diverge wildly, indicating that they cannot be described by any continuum stress model. We stabilize the packs by supplying an additional boundary, which makes the response much more regular. The response of the stabilized packs resembles what one would expect in a hyperstatic pack, despite the isostatic bulk. The expected stress rays characteristic of null stress behavior are not present. This suggests that isostatic packs do not need to obey a null stress condition. We argue that the rays may arise instead from more simple geometric considerations, such as preferred contact angles between beads.
\end{abstract}


\maketitle

\section{\label{Sec:Intro}Introduction}

The study of forces in static granular bead packs has a long tradition, owing to the rich panoply of observed phenomena. Extensive work has been done to study the distribution of of the inter-bead contact force strengths \cite{Liu1995,Blair2001,Silbert2002,vanEerd2007} and angles \cite{Ellenbroek2009}, the presence of load bearing arches \cite{Arevalo2006,Jenkins2011}, and the scaling of various properties such as pressure \cite{OHern2003, OHern2002}, effective shear modulus \cite{OHern2002, Wyart2008}, and length scales \cite{Silbert2005, Wyart2005}. The large assortment of interesting behaviors is due partly because the force response depends on the particular microscopic arrangement of the beads. Altering the geometric structure of the pack may have almost no effect at all, or it may cause huge changes to the bulk properties of the material. 

In this paper, we do not attempt to discuss how the forces will spread in an individual disordered pack. Rather, we attempt to describe the force propagation using a continuous, coarse-grained stress tensor by averaging over many different packs and finding the mean response. Such an approach is possible for hyperstatic granular packs \cite{Goldhirsch2002, Goldenberg2002}. However, it is not guaranteed to work for the isostatic packs we consider here.

Broadly speaking, the set of possible static granular packs can be divided into three types based on their stability. Hyperstatic systems are the most common. The number of contacts between beads is large, and if an external force is applied to the pack, forces can be generated across these contacts to stabilize the load. Hypostatic packs, on the other hand, do not have many contacts between beads. They tend to be less dense, and are unable to support some external loads. Such packs are characterized by the existence of so-called floppy modes, which are internal rearrangements of the bead positions which do not require any energy. Isostatic packs exist at the transition between these two types. They are stable in the sense that they can support any infinitesimal load provided. However, if even one contact is removed, they become hypostatic: a floppy mode is created, which typically causes a rearrangement of beads across the entire pack.

The stability of a pack can be estimated by comparing the number of beads in it with the number of contacts. In $d$-dimensional granular packs consisting of $N$ beads, static packs must satisfy $dN$ force balance equations. Without friction, the forces on any bead are either due to contacts with its neighbors or are applied externally. If the beads have, on average, $Z$ contacts each, then the total number of contacts in the pack is $n = NZ/2$. If this is less than $dN$, then the system is underdetermined. As discussed above, this happens in hypostatic packs. If $n>dN$, then there are more contacts than necessary. This is typically the case in hyperstatic packs. On the other hand, if $n=dN$, then often the pack is isostatic. In two dimensions, this means that the average contact number is $Z=4$. Because an isostatic pack has the same number of contacts as force balance equations, one can solve uniquely for all of the contact forces once the external forces acting on the system are known. Note that this isostatic contact number is necessary but not sufficient for the system to actually be isostatic; for a true test, one must check for the presence of floppy modes.

Isostatic packs behave quite differently than their hyperstatic brethren. Our goal is to find a continuous, coarse-grained stress tensor $\sigma_{\alpha\beta}$ at all points in the pack. In this continuum limit, the stress tensor should obey the continuity equations
\begin{equation}
\partial_\alpha\sigma_{\alpha\beta}=F_\beta
\label{Eqn:Continuity}
\end{equation}
where $\nkv F$ is the external force supplied to the system. These equations express static force equilibrium and, in two dimensions, produce two equations for the three components of the stress tensor. These two equations are not sufficient to determine the continuum stress field; a third piece of information is necessary. In a conventional elastic medium, this final equation is the constitutive stress-strain law, which reduces to the well known Hooke's Law for small deformations \cite[Section 4]{LandauTheoryOfElasticity}. This is not sensible in an isostatic medium where no contacts need to be compressed; as discussed above, on a microscopic level, force balance is actually sufficient to determine the response in any particular case. One proposal for a final continuum equation in the isostatic case is a linear relationship between the components of the stress tensor \cite{Tkachenko1999,Wittmer1996,Vanel2000,Cates1998}. 

For a quick idea of why the stresses should be linearly related in isostatic systems, imagine cutting out a small section from a large two dimensional isostatic pack. This will result in the creation of several floppy modes, due to the removal of the contacts on the sides. Now apply a stress to two opposite sides of this section. In an elastic material, there would be some resistance---energy could be stored in the resulting compression, and the sample would try to push back. However, the floppy modes allow the beads to move without using energy. The positions in two directions are restricted by the applied deformation, but the system is still free to move out in the other two directions. To confine the sample, we must also push on the section in these other directions, by an amount proportional to the force exerted on the sides. Thus in isostatic packs, we should have 
\begin{equation}
\sigma_{xx}=\mu\sigma_{xy}+\eta\sigma_{yy}
\label{Eqn:NullStressLaw}
\end{equation}
This argument does not depend on any packing history, and should hold for all isostatic packs. 

With the linear stress relationship established, the equations for each of the separate components of the stress tensor can be decoupled \cite{Tkachenko1999}. Taking derivatives of the continuity equations (\ref{Eqn:Continuity}) gives
\begin{align}
\partial_x\partial_y\sigma_{xx}+\partial_y^2\sigma_{xy} & =  0\\
\partial_x^2\sigma_{xy}+\partial_x\partial_y\sigma_yy & =  0
\end{align}
Plugging in the linear relationship from Equation~\ref{Eqn:NullStressLaw} allows us to eliminate $\sigma_{yy}$. This gives 
\begin{equation}
\big(\mu\partial_x\partial_y+\partial_y^2-\eta\partial_x^2\big)\sigma_{xy}  =  0
\end{equation}
which factors into
\begin{equation}
\big(\partial_y-c_+\partial_x\big)\big(\partial_y-c_-\partial_x\big)\sigma_{xy} = 0
\end{equation}
where 
\begin{equation}
c_\pm = \frac{-\mu\pm\sqrt{\mu^2+4\eta}}{2}
\end{equation}
Solving for $\sigma_{xx}$ and $\sigma_{yy}$ in a similar manner gives the same hyperbolic differential equation. Note that if $\mu=0$, then this becomes
\begin{equation}
\label{Eqn:WaveEqn}
(\partial_y^2-c^2\partial_x^2)\sigma_{\alpha\beta}=0.
\end{equation}
This is the wave equation, with the $y$ direction serving the role of time. Just as a flash of light will produce light cones extending out into spacetime along null lines, a point source of force will create lines of stress extending through the pack at angles determined by the ``speed of light,'' $c$. For this reason, we refer to the linear relationship as a ``null stress'' condition. In keeping with this analogy, we will refer to the lines of stress as ``light cones.'' In packs with left-right symmetry, the $\mu=0$ condition should hold. This is the case in all of the packs we examine. Even if $\mu\ne 0$, however, we can still rotate into a frame where $\mu$ vanishes and recover the wave equation. In this case, the light cones will still exist, but will be bisected by some direction other than the vertical.

\begin{figure}
\begin{tabular}{cc}
\resizebox{35mm}{!}{\includegraphics{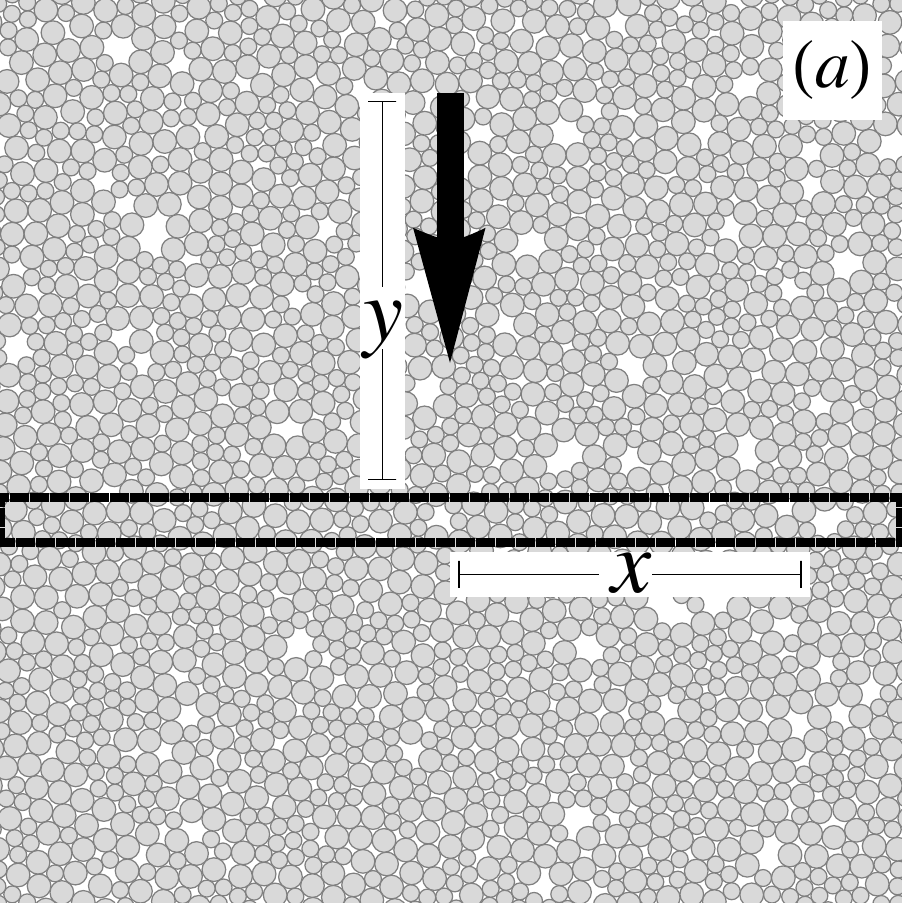}} & 
\resizebox{45mm}{!}{\includegraphics{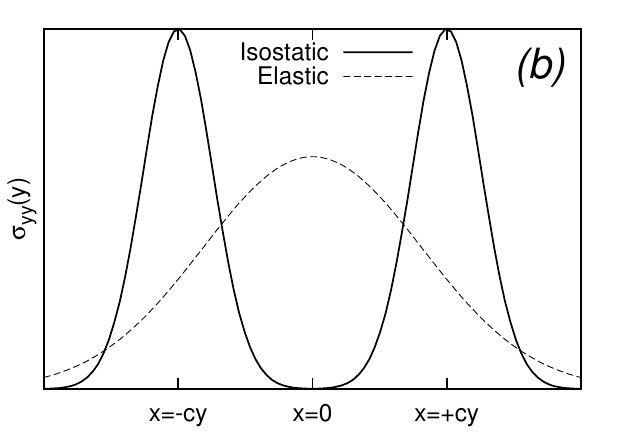}}
\end{tabular}
\caption{\label{Fig:CompareStressCartoon}Cartoons of $\sigma_{yy}$ as a function of $x$, a distance $y$ below the forcing point. In (a), we show the setup: a point force is supplied to the pack, and we look at $\sigma_{yy}$ in a region a distance $y$ below it. In (b), we see the expected stress in this region for both elastic and isostatic media. The elastic solid has a large central maximum below the applied force, whereas the isostatic pack has two peaks, corresponding to the positions of the light cones.}
\end{figure}

This hyperbolic behavior is not seen in elastic media or hyperstatic granular packs. In those systems, there is a single maximum below the location of the forcing point, whereas in a hyperbolic system there is a minimum with two peaks to either side, as depicted in Figure~\ref{Fig:CompareStressCartoon}. There are some models where similar rays of force are expected in hyperstatic packs \cite{Claudin1998,Bouchaud2001}. In these models, there is a crossover region where at a certain depth the positional disorder in the system causes the light cones to merge into a central maximum following elliptic rather than hyperbolic behavior. This is different than what we are describing here, and we do not expect such a central maximum \cite{Blumenfeld2004}. In Section~\ref{Sec:IsotropicPacks}, we will briefly discuss how the noise caused by disorder obeys different statistics in isostatic packs.

To test this null stress theory, we can supply a small force to a bead and see if the stress is, on average, concentrated into light cones. Unfortunately, experimental tests have so far proven inconclusive \cite{Moukarzel2004, Geng2001}, due to various difficulties such as creating packs which satisfy the isostatic conditions, measuring the forces in the bulk, eliminating friction, and getting enough data to provide a meaningful average. In the absence of real-world verification, we turn to simulation. According to Huygens's principle, in $n+1$ dimensions, the Green's function for the wave equation has a long tail if $n$ is even. This makes signals harder to distinguish than they would be if $n$ were odd \cite[Chapter 6]{GarabedianPartialDifferentialEquations}. Therefore we work in two dimensions, where the light cones are easier to see. The beads that we use are frictionless, and interact only with their neighbors via radial harmonic contact forces. We will allow these contact forces to be tensile. In the next section, we will argue that the particular choice of force law should not have a large effect on the measurements we are concerned with.

We begin by providing an overview of the systems in question in Section~\ref{Sec:Background}. In Section~\ref{Sec:SequentialPacks}, we describe previous simulations which appear to verify the null stress law, and highlight a broken symmetry inherent in the construction of these systems. We attempt to make a more general system in Section~\ref{Sec:IsotropicPacks}, by creating isotropic isostatic packs, and note disagreements with the null stress law: not only do we see the response fail to produce light cones, but we find that there is no continuum stress field at all. In Section~\ref{Sec:HardTop}, we show that it is possible to restore continuum behavior, but produce elastic-like responses rather than light cones. In Section~\ref{Sec:ContactAngles}, we argue that the light cones found in previous simulations may have been caused by the contact angle distribution of the packs they used, rather than a null stress effect. We then offer concluding remarks in Section~\ref{Sec:Conclusion}.

\section{\label{Sec:Background}Background}

We describe here the methods we use to analyze the packs after they have been created. We will work in the regime of linear response, where the magnitudes of the forces we supply are infinitesimal. Since the system is jammed, the positions of the beads do not change in response to these forces, and they can be balanced via contact forces with other beads. For the system as a whole, we can write this as
\begin{equation}
M\nkv f=\nkv F
\label{Eqn:ForceBalance}
\end{equation}
where $\nkv f$ is a vector containing the values of the contact forces for each  of the $n$ pairs of touching beads, $\nkv F$ is a vector of length $2N$ containing the $x$ and $y$ components of the external force on each bead, and $M$ is a $2N\times n$ contact matrix. Typically some sort of hard floor will support the pack. This floor is in contact with beads, but we do not require the net force on it to be zero. The floor can thus balance any net force applied to the bulk of the pack. 

Since the system is isostatic, $M$ is square, and for any given set of body forces one can uniquely solve for the network of inter-bead contact forces which keep the system stable. Furthermore, $M$ depends only on geometry; once the contact angles are known, nothing else is required to solve the problem. Each row of $M$ represents either the $x$ or $y$ components of the contact forces exerted on one bead. The row has one non-zero entry for each contact that bead has. The value of this entry is given by the unit vector pointing from the row's bead to the other bead it is in contact with, projected onto whatever direction the row corresponds to. This gives each column of $M$ a simple structure. Say that column $\alpha$ is due to the contact between beads $i$ and $j$, and $\nkv n_{ij}$ is the unit vector pointing from $i$ to $j$. The column will have exactly four nonzero entries: in the rows corresponding to bead $i$'s $x$ and $y$ components, it will have $(n_{ij})_x$ and $(n_{ij})_y$, and for the rows corresponding to bead $j$, it will have $(n_{ji})_x$ and $(n_{ji})_y$. Furthermore, $\nkv n_{ij}=-\nkv n_{ji}$, and $|\nkv n_{ij}|=1$, so each column depends on only one unique piece of information. Note that if the contact is between a bead and the hard floor rather than between two beads, then the column will have only two nonzero entries. In two dimensional isostatic systems, the average number of contacts per bead is four, so as the system size grows, only a vanishingly small fraction of the entries are nonzero. $M$ is evidently quite sparse. 

If, for some $\nkv F$, there is no $\nkv f$ which satisfies Equation~\ref{Eqn:ForceBalance}, then the system is not jammed: no set of contact forces can prevent at least one bead from experiencing a net force and shifting in response. However, as long as our system is isostatic, this will not happen. The positions of the beads remain fixed, and so the contact angles can never change; $M$ is therefore constant. 

Given an $\nkv F$, some of the elements of $\nkv f$ may be negative, indicating tensile forces. Therefore, this cannot strictly represent a hard sphere system where all contacts are purely repulsive. However, Tkachenko and Witten \cite{Tkachenko2000} have developed an ``adaptive network'' technique which forces all of the contacts in an isostatic system to be compressive. They do this by making slight changes to the packing geometry so that under a compressive load, there are no tensile forces. They discovered that the response in their system followed the null stress prediction both before and after this procedure, indicating that the existence of light cones does not depend on any restrictions, or lack of restrictions, on the sign of the contact forces. Thus for simplicity, we simply allow tensile forces in our simulations. There are many examples of physical systems where this is a realistic model. For example, if neighboring beads are slightly wet, they can form liquid capillary bridges which hold them together \cite{Fingerle2008}, and colloids often have attractive interactions due to, for example, depletion \cite{Eckert2002}. Often such systems have a yield point, where a large enough tensile force will split the beads apart. However, in the linear response regime appropriate for the small forces in question, the contact force magnitude should be small enough that this is not a concern.

Instead of using the contact matrix $M$ described above, the linear response of the system can also be measured using a dynamical matrix $\mathcal D$ \cite{Ellenbroek2009}. While the contact matrix depends only on the angles between beads, the dynamical matrix assumes a spring constant $k$ between each of the connected pairs. With this spring constant, the energy of the system is quadratic in the small displacements $\nkv u$ each bead has from its equilibrium position: 
\begin{equation}
\label{Eqn:DynamicalMatrix}
\mathcal E = \frac k2 \nkv u^T \mathcal D \nkv u
\end{equation}
This $\mathcal D$ matrix must be symmetric, and it follows that the force the springs exert on each of the beads is $- k \mathcal D \nkv u$. Thus in equilibrium we can relate the body force on each bead to its displacement via $\nkv F = k \mathcal D \nkv u$. This is an equivalent way of finding the response. To translate between the two, note that the spring energy can also be written in terms of the contact forces. Then
\begin{align}
\label{Eqn:ConvertMtoD}
\mathcal E = \frac{1}{2k}\nkv f^T\nkv f & =  \frac{1}{2k}\big(\nkv F^T(M^T)^{-1}\big)\big(M^{-1}\nkv F\big)\\ \notag
& =  \frac{1}{2k}(k \nkv u^T \mathcal D^T)(M^T)^{-1}(M^{-1})k\mathcal D\nkv u \\ \notag
& =  \frac{k}{2}\nkv u^T\big(\mathcal D (MM^T)^{-1}\big)\mathcal D\nkv u \nonumber
\end{align}
Since $\mathcal D$ is symmetric and this must be true for all $\nkv u$, comparison with Equation~\ref{Eqn:DynamicalMatrix} shows that the dynamical matrix can be written down solely in terms of the contact angles: $\mathcal D=MM^T$. However, whereas the contact matrix only depended on these angles and not the functional form of the interaction between beads, by using the dynamical matrix we are implicitly assuming a harmonic interaction. This choice of a linear spring force between contacts is not realistic for some physical systems. The contact between smooth, solid bodies is more accurately described by Hertz's Law \cite{JohnsonContactMechanics}, in which the the force varies as the 3/2 power of the compression rather than linearly. However, the properties we study should depend only on the isostatic nature of the contact network, and not on the force law of the contacts. Previous simulations have verified that the density of states and other features at the isostatic point are unchanged if one uses Hertzian rather than harmonic contact forces \cite{Silbert2005}. Indeed, all of our systems have zero pressure, so any particular choice of overlap potential has no chance to cause an effect. Once the final bead configuration is reached, all analysis for the response can be done in terms of the contact matrix $M$, which depends only on the angles between beads in contact, and not any displacements due to a particular force law.

Note that Equation~\ref{Eqn:ConvertMtoD} required the use of $M^{-1}$, which does not exist if the system is not isostatic. However, even if $M$ is not invertible, it remains true that $\mathcal D=MM^T$. The proof is more tedious in this case, and as such has been relegated to Appendix~\ref{Sec:DMMTProof}.

If each of the beads has some mass $m$, then $m\ddot{\nkv u} + k \mathcal D \nkv u = 0$. Thus the spectrum of $\mathcal D$ gives the vibrational modes of the system: the squared normal more frequencies $\omega_i^2$ are given by the eigenvalues of $k\mathcal D/m$. We can use these frequencies to measure the density of states, $D(\omega)$, defined as 
\[
\label{Eqn:DensityOfStates}
D(\omega) = \lim_{\Delta\omega\rightarrow 0}\frac{\textrm{(\# of $\omega_i$ such that $\omega\le\omega_i<\omega+\Delta\omega$)}}{2N\Delta\omega}
\]
If $\mathcal D$ is singular, then some $\omega_i$ will be 0. Their corresponding eigenvectors are the floppy modes of the system.

\begin{figure}
\resizebox{80mm}{!}{\includegraphics{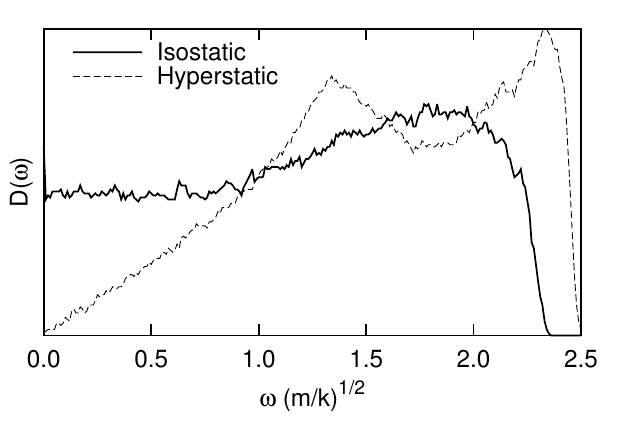}}
\caption{\label{Fig:CompareDensityOfStates}The density of states for an isostatic pack has a plateau at low frequencies. A hyperstatic pack, on the other hand, has a density of states which goes to zero at small frequencies. In this it resembles an elastic solid, where $D(\omega)\sim\omega^{d-1}$, with $d$ the spatial dimension. The isostatic pack here was created using the method described in Section~\ref{Subsec:PackCreation}, and the hyperstatic pack was found by treating all of the nearest-neighbor beads in this same pack as being in contact. This gives it an average of $Z=5.7$ contacts per bead.}
\end{figure}

In $d$-dimensional hyperstatic packs, $D(\omega)\sim\omega^{d-1}$ for small $\omega$. However, the density of states of an isostatic pack has a large nonzero plateau at low frequencies, as shown in Figure~\ref{Fig:CompareDensityOfStates}. If the pack is only slightly hyperstatic, in the sense that the contact number is only a little larger than the isostatic value of four, then the density of states has been shown to grow linearly from zero until some $\omega^*$, above which it will exhibit the plateau expected in an isostatic pack \cite{Wyart2005,Silbert2005}. An isostatic pack will behave differently from the hyperstatic one below this $\omega^*$. It is these low frequencies which should be most important for static force propagation. 

In a hyperstatic system $M$ may not be square, so it cannot be inverted to give $\nkv f$. To find the equilibrium forces in this case, we treat each contact like a harmonic spring, as discussed above. The contact force vector is then the $\nkv f$ which satisfies $M\nkv f=\nkv F$ while minimizing the energy $\frac{1}{2k}\nkv f^T\nkv f$ from Equation~\ref{Eqn:ConvertMtoD}. 

After the contact forces are known, the stress can be computed using an established local coarse graining procedure \cite{Goldhirsch2002,Ellenbroek2009}, which sums up the contributions from all pairs $(i,j)$ of contacting beads:
\begin{equation}
\label{Eqn:StressFromContactForces}
\sigma_{\alpha\beta}(\nkv x) = \sum_{(i,j)} f_{ij}R_{ij}(n_{ij})_\alpha (n_{ij})_\beta\int_0^1 ds \Phi(\nkv x-\nkv x_j+s \nkv R_{ij}).
\end{equation}
Here $\nkv R_{ij}$ is the vector pointing from bead $i$ to bead $j$, whose unit vector is $\nkv n_{ij}$. $\nkv x_j$ is the position of bead $j$, and $\Phi$ is a non-negative coarse-graining function centered at the origin whose integral over all space is 1. Often $\Phi$ is chosen to be a Gaussian or a smoothed Heaviside function. For simplicity, we will use
\begin{equation}
\label{Eqn:CoarseGrainDefinition}
\Phi(\nkv r) = \left\{\begin{array}{ll} \frac{1}{\pi w^2}; &  r < w\\ 0; & \textrm{otherwise}\end{array}\right.
\end{equation}
where the width $w$ will often be chosen to be on the order of a bead diameter.

We would like to measure the response in a way that takes into account the direction and magnitude of the source force, $\nkv F$. The vector $(\nkh F \cdot \sigma)$ has to be proportional to the strength of the applied force. Using Equation~\ref{Eqn:StressFromContactForces}, the constant of proportionality is
\begin{equation}
\label{Eqn:ProportionalStress}
\frac{\nkh F\cdot\sigma(\nkv x)}{|\nkv F|} = \sum_{(i,j)}\Big(\int ds \Phi(\nkv x-\nkv x_j+s\nkv R_{ij})\Big) \nkv G_{ij}(\nkv x_j)
\end{equation}
where
\begin{equation}
\label{Eqn:DefineG}
\nkv G_{ij}(\nkv x_j)=\frac{f_{ij}}{|\nkv F|}R_{ij}\nkv n_{ij} (\nkv n_{ij}\cdot \nkh F)
\end{equation}
The response function $\nkv G$ contains the information which gives the relationship between the contact forces and the applied external force. It can easily be computed at the location of any contact without reference to coarse-graining integrals. This response is similar to the one used by Head, Tkachenko, and Witten \cite{Head2001}, but it differs in that it is more directly related to the stress. This $\nkv G$ has units of length, given by the factor of $R_{ij}$ which tells the distance between the beads in question. In the following, we will always report $\nkv G$ in units of the average diameter of the beads in the pack.

\begin{figure}
\begin{tabular}{cc}
\resizebox{40mm}{!}{\includegraphics{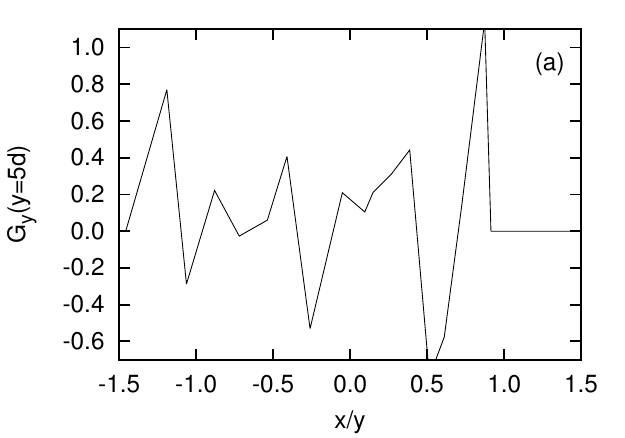}} & 
\resizebox{40mm}{!}{\includegraphics{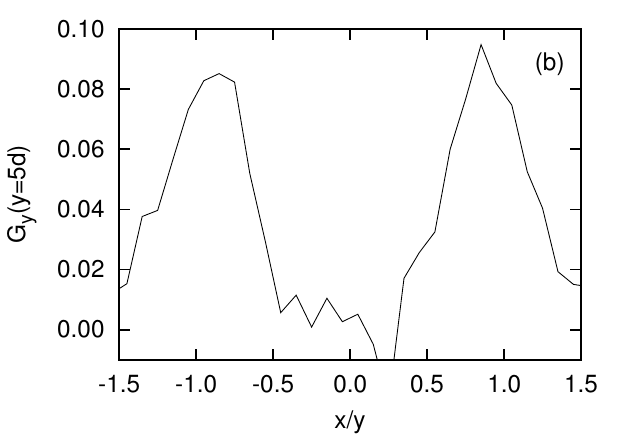}}
\end{tabular}
\caption{\label{Fig:AveragingResponses}The response, computed using the techniques described in Section~\ref{Sec:Background}, to a point force (a) in a single pack with 512 beads and (b) averaged over 2970 different realizations. The packs were generated using the sequential method of Section~\ref{Sec:SequentialPacks}, and for each different pack, several realizations were generated by applying the point force to different beads near the top of the pack. The response for each realization was measured along a line five bead-diameters below that point. We choose five bead-diameters here, and elsewhere in this paper, but the light cones are visible at all depths below the applied force. This is shown clearly by Head et. al. \cite{Head2001} and Moukarzel \cite{Moukarzel2002}. We use five bead-diameters because it provides a balance between being deep enough in the pack that the light cones have had room to spread apart enough to be resolveable, and being close enough to the applied force that excessive averaging is not required. Even though the response for the single pack varied wildly throughout, on average the large deviations in the middle cancel, leaving stress only in the region where $x\approx y$.}
\end{figure}

Many of our results will be presented in terms of $\nkv G$. It must be averaged over several different systems to find a continuum response. Each instance is individually quite noisy, but across several realizations the average will be smoother, as shown in Figure~\ref{Fig:AveragingResponses}.

\begin{figure}
\resizebox{60mm}{!}{\includegraphics{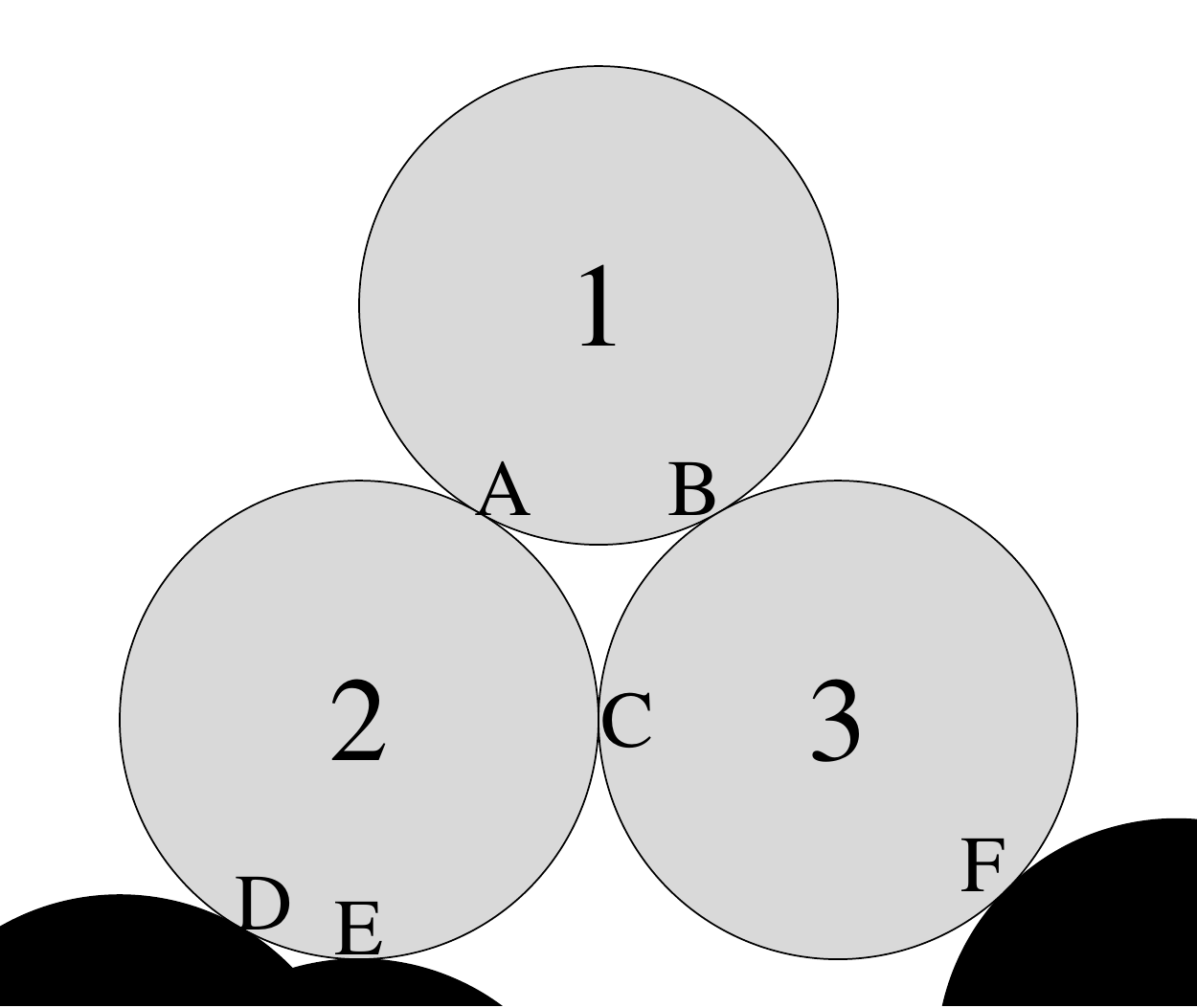}}
\caption{\label{Fig:SampleComputation}A very small sample pack to illustrate the method of computing the response described in Section~\ref{Sec:Background}. Three beads of radius $R$, numbered 1-3, form an equilateral triangle. They sit on a rough floor, so that bead 3 is touching the ground at a 45$^\circ$ angle, and bead 2 is touching the ground in two places: one directly below the center, and another 30$^\circ$ to the side. The six contacts are labeled A-F. Since there are six contacts and three beads, the system is isostatic.}
\end{figure}

As a concrete example of how this works in practice, consider the small example pack depicted in Figure~\ref{Fig:SampleComputation}. We will write $M$ so that the $x$ and $y$ components of force balance for bead $i$ correspond to rows $2i-1$ and $2i$ respectively. The columns will represent contacts A-F, in order. Then we have
\begin{equation}
M=\frac 12 \left(\begin{array}{cccccc}
-1 & 1 & 0 & 0 & 0 & 0\\
-\sqrt{3} & -\sqrt{3} & 0 & 0 & 0 & 0\\
1 & 0 & 2 & -1 & 0 & 0\\
\sqrt{3} & 0 & 0 & -\sqrt{3} & -2 & 0\\
0 & -1 & -2 & 0 & 0 & \sqrt{2}\\
0 & \sqrt{3} & 0 & 0 & 0 & -\sqrt{2}
\end{array}\right)
\label{Eqn:SampleM}
\end{equation}
To supply a unit force down on bead 1, we use $\nkv F=(0,-1,0,0,0,0)^T$. Since the system is isostatic, we can simply solve Equation~\ref{Eqn:ForceBalance} to get the contact force vector, 
\[
\nkv f = \Big(\frac{1}{\sqrt{3}},\frac{1}{\sqrt{3}},\frac{3-\sqrt{3}}{6},1,\frac{1-\sqrt{3}}{2},\frac{1}{\sqrt{2}}\Big)^T.
\]
If the system had been hyperstatic, then we would instead solve the linear least squares problem \cite{Haskell1981} of minimizing $|\nkv f|^2$ subject to the constraints $M\nkv f=\nkv F$. Routines to solve this problem are available in many scientific programming packages. Once the contact force vector $\nkv f$ has been computed, the response $\nkv G$ can be found for each contact. For example, the response for contact A is located at $(x,y) = (-R/2,-R\sqrt{3}/2)$ relative to the applied force, and takes the value $\nkv G=-f_A(R+R)\nkv n_{12}(n_{12})_y = (-1/4,-\sqrt{3}/4)$, in units of the beads' diameters. The response at the other contacts can be computed in a similar manner. 

\section{\label{Sec:SequentialPacks}Sequential Packs}

Previous simulations have shown the presence of light cones in periodic two dimensional packs built up by adding variably sized disks one by one onto an irregular floor \cite{Tkachenko1999, Tkachenko2000, Head2001}. The results indicated a speed of light consistent with 1, meaning that the rays of force propagated out at 45$^\circ$ angles below a supplied point force. Examples of these light cones were shown in Figure~\ref{Fig:AveragingResponses}. Such sequential packs are formed by taking beads one at a time and placing them each at the lowest point in the system where they will have exactly two contacts. Using this construction, the system has the isostatic contact number at each step, and the process can easily be repeated until the desired system size is reached. To avoid crystallization, the bead radii are chosen from a polydisperse distribution. Unless noted otherwise, the packs in this paper are bidisperse, with a radius ratio of 1.4 : 1. In some other similar situations, the beads have a weight, and gravity plays a role in the deposition process. This leads to the formation of local structures such as bridges \cite{Pugnaloni2004}. However, here there is no gravity. The lowest point in the pack is determined by the starting location of the floor, and there are no forces on any of the beads until the point force is supplied. Only when this happens are there any nonzero contact forces. While in an elastic solid there would be a nonzero response both above and below the applied force, in these isostatic packs the contact forces only appear below the supplied force \cite{Head2001}, and propagate down toward the floor in the manner indicated in Figure~\ref{Fig:AveragingResponses}.

\begin{figure}
\resizebox{80mm}{!}{\includegraphics{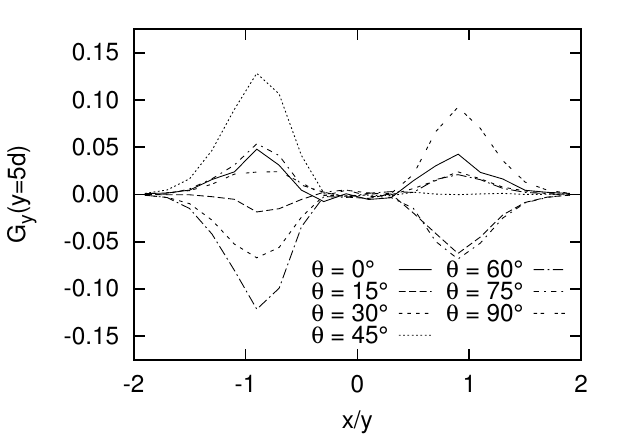}}
\caption{\label{Fig:PullAtAngle}The response five bead-diameters below a point force in a two-dimensional sequential pack. $G_y$, measured in units of bead-diameters, is the vertical response discussed in Section~\ref{Sec:Background}, and $\theta$ is the angle of the applied force relative to the $-y$ direction. This response was calculated by averaging over 4500 realizations. Changing the direction of the applied force does not change the location of the peaks; only the relative magnitudes differ.}
\end{figure}

Discovering what determines the direction of the resulting light cones was the motivating factor in this work: why do they go ``down''? There are three obvious ways in which symmetry can be broken to allow for this. First, the direction of the applied point force provides an option for a preferred direction. However, this is easily shown not to be the deciding factor. No matter what angle the point force is directed in, the locations of the light cones do not change; the only difference is their relative magnitudes, as shown in Figure~\ref{Fig:PullAtAngle}. Second, the packing history contains a clear distinction between up and down: new beads are always placed above the ones already there. However, the argument for the null stress law makes no reference to packing history, so changing this anisotropic construction should not affect the light cones. Third, the boundary conditions clearly provide a distinction between the unconstrained top and the hard floor.

To separate the effects of the anisotropic boundaries from those of the anisotropic packing history, we construct packs in an isotropic manner, and then later break the symmetry by adding a floor. Because the boundary conditions are the same in both cases, the above argument might lead one to expect the response to be unchanged. However, as detailed in Section~\ref{Sec:IsotropicPacks}, we do not see light cones in this isotropic case!

\section{\label{Sec:IsotropicPacks}Isotropic packs}

In this section we test whether an isotropic pack can have light cone force propagation like the sequential packs of the last section. Certainly no light cones are expected if the system gives no indication of which direction they can point. As noted above, we may define such a direction by adopting one feature of the sequential pack, namely the constraining floor. We first describe our procedure for creating an isotropic pack in a periodic box. Next we show how we modify the system to create a floor constraint. We then discuss the effects of this modification and argue that the isostatic character of the pack is not essentially altered. Finally, we discuss the peculiar behavior of the force response using different boundary modifications.

\subsection{\label{Subsec:PackCreation}Pack creation}

We follow an established method to create isostatic packs with no preferred direction \cite{OHern2003}. We start with a periodic $L\times L$ box filled with small beads at random locations. The bead radii are chosen according to whatever polydispersity we please such that the system's density $\phi=\sum_i(\pi R_i^2)/L^2$ is well below the critical density $\phi_c$ where the system jams. Though $\phi_c$ varies from pack to pack, in two dimensional systems it is typically larger than 0.82 \cite{OHern2003}. 

The beads are given a pairwise interaction potential which is harmonic if the two beads $i$ and $j$ are overlapping, and zero otherwise: 
\[
V_{ij}=\left\{\begin{array}{ll} \varepsilon \big(1 - \frac{r_{ij}}{R_{ij}}\big)^2 & r_{ij}<R_{ij} \\ 0 & \textrm{else}\end{array}\right.
\]
where $r_{ij}$ is the distance between beads $i$ and $j$, $R_{ij}=R_i+R_j$ is the sum of their radii, and $\varepsilon$ is a constant that sets the energy scale. With this potential, we use a conjugate gradient algorithm to minimize the energy of the system by rearranging the bead positions. Since our system is fairly dilute at this point, it is possible to find a configuration, close to the original, where the energy is zero.

Next we swell all of the beads by some percentage so that the packing fraction increases by a small amount $\Delta\phi$, and again perform the conjugate gradient minimization. We do this quasistatically so that the bead rearrangements are minimal; typical $\Delta\phi$ depend on the system size, but are on the order of $10^{-4}$ to $10^{-10}$.

Eventually the beads will be large enough that no local rearrangement can remove all bead overlaps, and the conjugate gradient algorithm will give a system with some finite pressure. It has been shown empirically that at this transition, the system is isostatic \cite{OHern2003}. In practice, due to our finite step size $\Delta\phi$, minor adjustments need to be made to get a fully isostatic pack. We first count the number of contacts $n$ relative to the number of beads $N$. Note that around 5\% of the beads will have no contacts. These are called floaters or rattlers, and we do not include them in our count of $N$; when the final configuration is reached, we actually remove them from the system altogether, since they cannot contribute to the force network. If $n=2N$, then the system has the isostatic contact number and no adjustments are necessary. If our system is hyperstatic by a few contacts, then we decrease $\Delta\phi$ by a factor of two and reverse the sign, so that the beads shrink slightly instead of swelling. We continue to change the packing fraction until the system is isostatic, or it becomes hypostatic. In the latter case, we again take $\Delta\phi\rightarrow-\Delta\phi/2$. This is repeated until we either get the isostatic contact number, or $\Delta\phi$ gets too small and we give up. 

Once the system has the isostatic contact number, we can check to see if it is truly isostatic by counting its floppy modes. There are necessarily two such modes of $\mathcal D$ with zero eigenvalue, corresponding to the two different uniform translations available under the periodic boundary conditions, but these will disappear when the floor is added and can thus be ignored here.

In these systems, the four directions given by the periodic boundaries are of course the same, but this construction might allow some other direction to be different. Nevertheless, we refer to these systems as isotropic, in light of the fact that they are explicitly not anisotropic in the directions relevant to the sequential packs of Section~\ref{Sec:SequentialPacks}: there is no difference between up-down and left-right.

Recent work suggests that because our pack creation algorithm is driven purely by volume-changing processes, it will be stable under uniform compression but not under volume-preserving shear forces. This can result in systems with a negative shear modulus \cite{DagoisBohy2012}. Furthermore, due to the extra degree of freedom allowed by boundary distortions, the number of contacts required for a finite system to jam may not be the same as the isostatic number discussed earlier. In this case, rather than using the lack of floppy modes as the determining characteristic for jamming, the existence of nonzero bulk and shear moduli can be used instead \cite{Goodrich2012}. However, all of the systems that we study are modified versions of these periodic packs, where at least one of the boundaries will be rigidly fixed in place, removing its degree of freedom. We find that counting floppy modes is sufficient to make our pack stable under the forces we apply. 

\begin{figure}
\begin{tabular}{cc}
\resizebox{40mm}{!}{\includegraphics{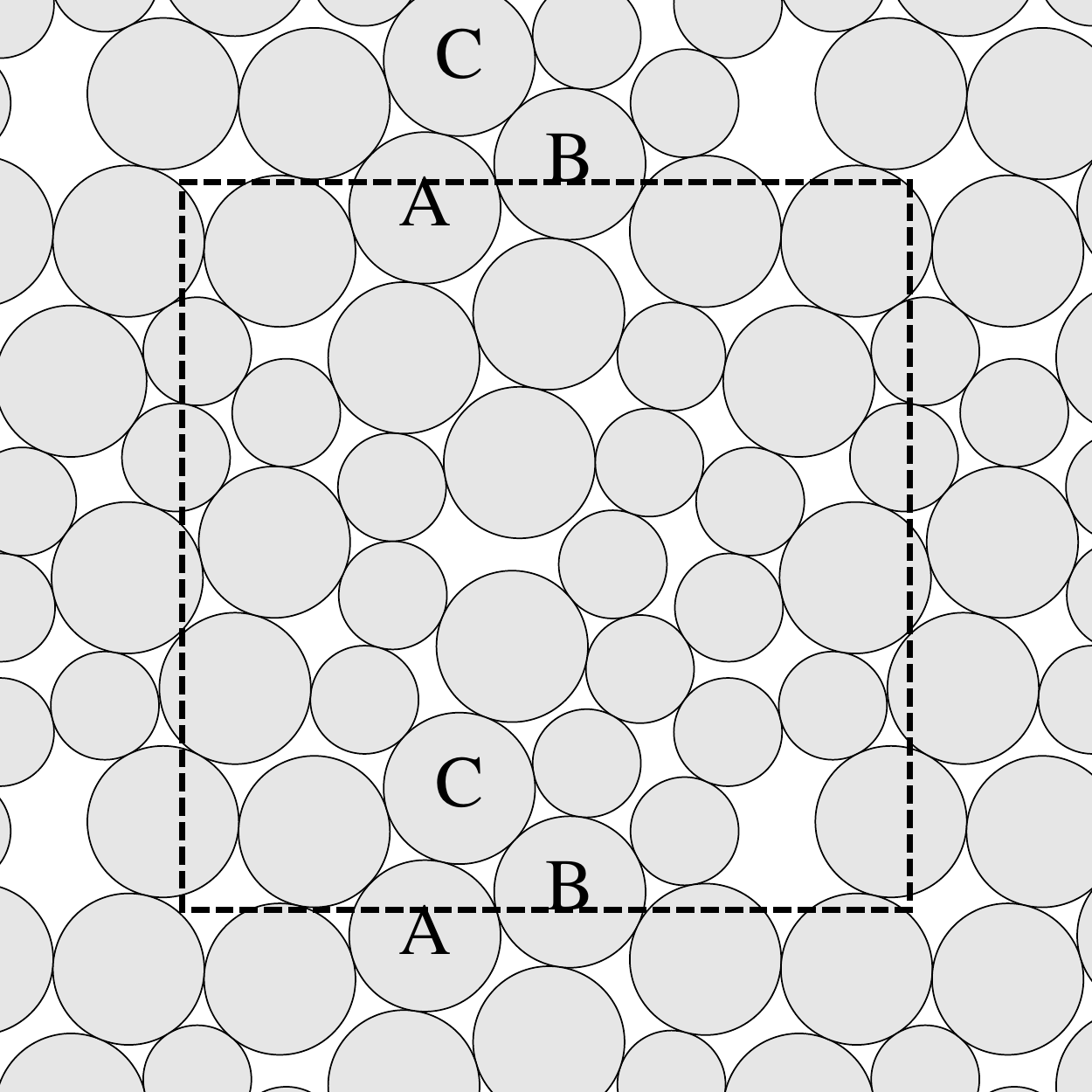}} & 
\resizebox{40mm}{!}{\includegraphics{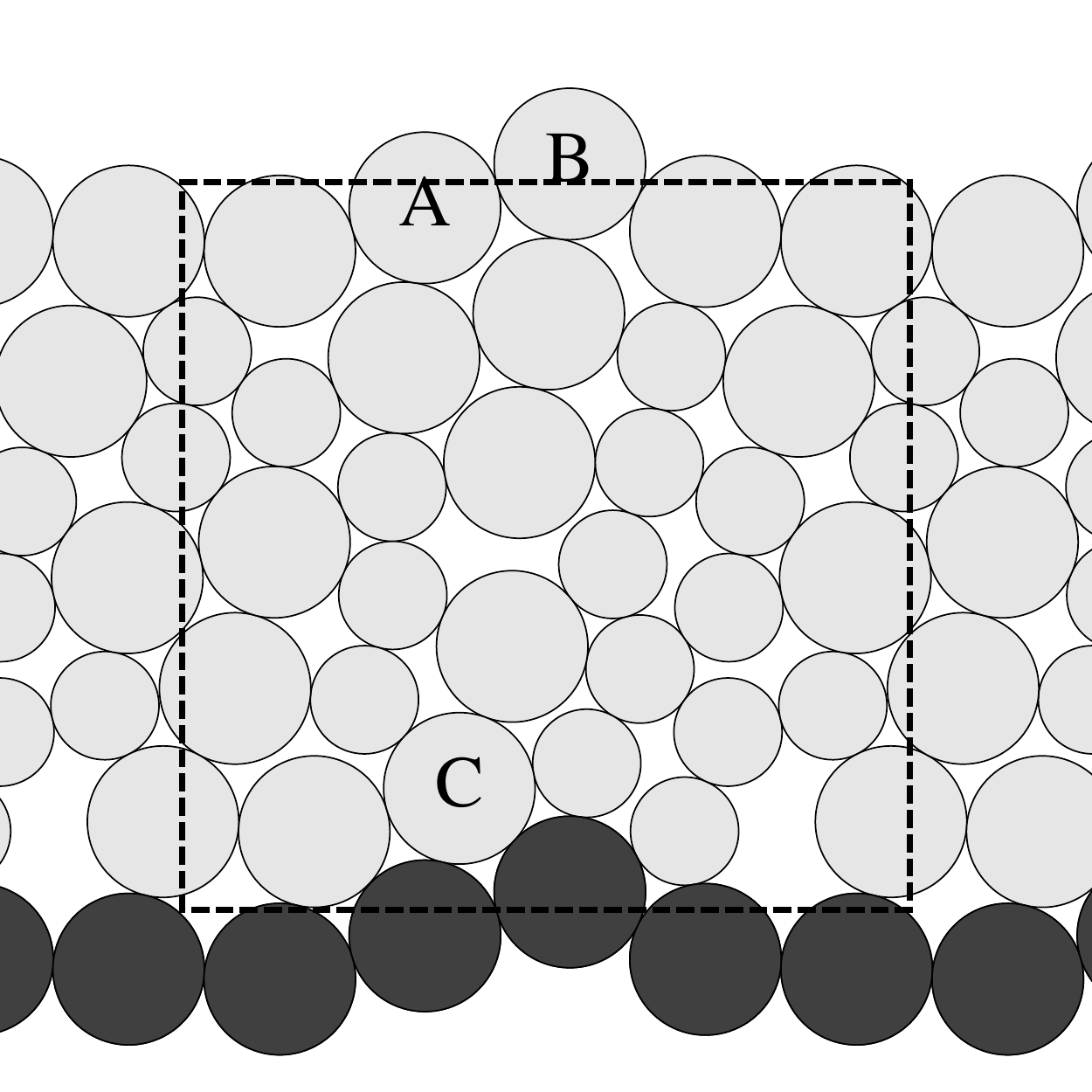}}
\end{tabular}
\caption{\label{Fig:AddFloor}On the left, a small pack created using the method described in Section~\ref{Sec:IsotropicPacks}, with a dashed box indicating the periodic boundaries. Three beads have been labeled for comparison with the image on the right, which denotes the same pack with a floor. Beads which cross the horizontal boundary (e.g. A and B) are treated as being on top of the pack. This means that the force balance equations for these beads do not include the contact with bead C. However, the force balance equation for C remains unchanged; what was formerly the contact from C to A is now a contact from C to the floor. In this way, both the total number of beads in the pack and the total number of contact forces remains unchanged.}
\end{figure}

To break the symmetry, we now create a bumpy floor to resemble the sequential case. This is done by taking some arbitrary horizontal line across the pack, and designating all beads it crosses to be on the boundary. Each boundary bead serves both as a floor for the beads above it to rest on, and also as a bead sitting on the top of the pack. This is accomplished by keeping the force balance equations the same as they would be otherwise, except that for beads on the boundaries, contacts with non-boundary beads above them are not included. This is explained in more detail in Figure~\ref{Fig:AddFloor}.

This modified system has the same number of beads and contacts as the old one, and thus still globally satisfies the isostatic contact number. However, there are now fewer force balance equations on the bottom, since none of them are used for the floor. Because the number of contacts did not decrease, there is a locally hyperstatic region of the pack here. On the top, we have a corresponding hypostatic region. This causes two problems. First, the existence of a hyperstatic region means that the contact matrix $M$ has at least one null vector involving the contacts on the bottom of the pack. This is a nonzero pattern of contact forces which sum to zero net force on every bead. Thus there is no unique solution to our equation $M\nkv f=\nkv F$, as arbitrary multiples of these null vectors can be added to $\nkv f$ to give the same $\nkv F$. This problem can be solved by finding the $\nkv f$ which minimizes the compressive energy. As discussed in Section~\ref{Sec:Background}, this is done by choosing the $\nkv f$ for which $|\nkv f|^2$ is a minimum.

\begin{figure}
\begin{tabular}{cc}
\resizebox{40mm}{!}{\includegraphics{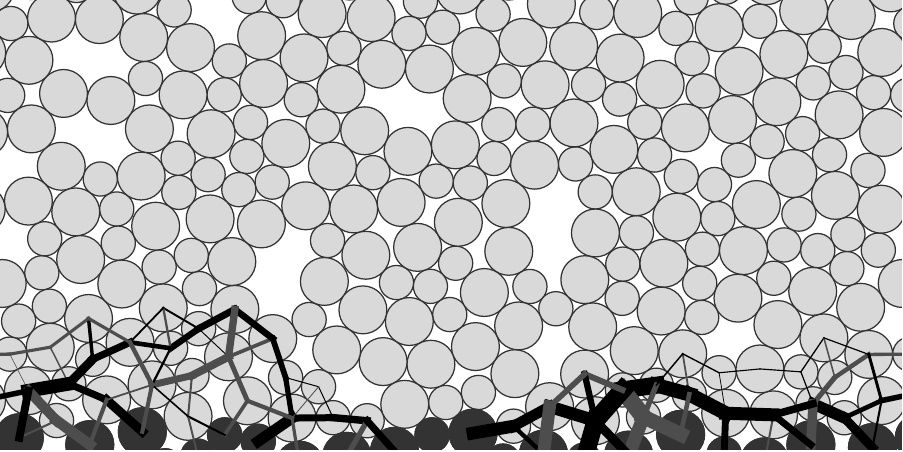}} &
\resizebox{40mm}{!}{\includegraphics{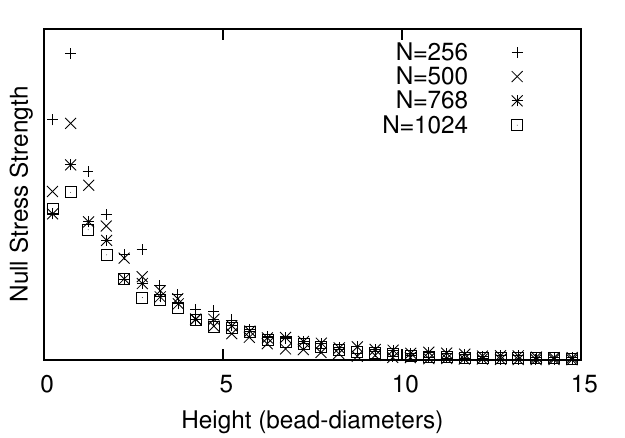}}
\end{tabular}
\caption{\label{Fig:HyperstaticNullVectors}On the left, an image of one of the hyperstatic null vectors of the contact matrix $M$. $M$ becomes singular when a hard floor is added using the method shown in Figure~\ref{Fig:AddFloor}. The lines between beads represent contact forces, with color denoting sign and thickness denoting magnitude. Even if some multiple of this nonzero pattern of contact forces is put on the beads, there will still be zero net force on all beads. On the right, we see that the null vectors are localized near the boundary. For several different pack sizes, we look at the null vectors formed when the hard floor is added. Each null vector is scaled so that its L2 norm is 1, and then the average magnitude of all contact forces at a particular height is computed. The average is taken over all null vectors created in each of several different floor configurations. The different floor configurations were chosen by putting the horizontal boundaries that determine them at several heights, separated by a distance of two bead-diameters. No matter the size of the pack, the strength of the null space decayed following an exponential whose decay length was between two and three bead-diameters.}
\end{figure}

\begin{figure}
\begin{tabular}{cc}
\resizebox{40mm}{!}{\includegraphics{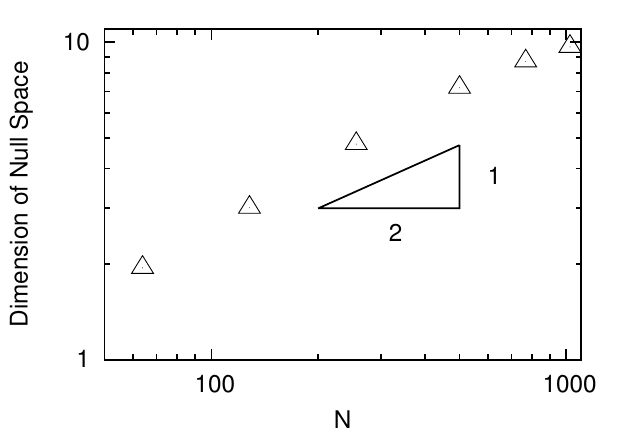}} &
\resizebox{40mm}{!}{\includegraphics{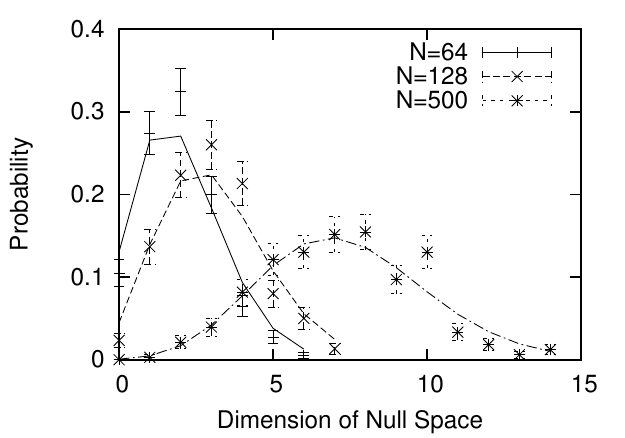}}
\end{tabular}
\caption{\label{Fig:NumberNullVectors}When a hard floor is added to the system using the method shown in Figure~\ref{Fig:AddFloor}, the contact matrix $M$ becomes singular. In (a), we show that the average number of null vectors of $M$ scales with the linear size of the system; that is, with the square root of the number of beads $N$. Thus as the system size is increased, a vanishingly small percentage of the modes of $M$ will have a zero eigenvalue. In (b), we show the probability that a certain null space dimension was measured, given the size of the system. The lines show a Poisson distribution fitted using a mean equal to the average null space dimension for that system size.}
\end{figure}

Though we give a prescription for choosing a particular $\nkv f$, we do not actually expect this choice to matter in the bulk of the pack where our measurements are made. Figure~\ref{Fig:HyperstaticNullVectors} shows that the magnitudes of the contact forces in the hyperstatic null vectors decrease exponentially with distance from the boundary, with a decay exponent on the order of a few bead-diameters. Thus while multiples of these null vectors can be added to any solution, only the contact forces near the floor are affected. Moreover, the number of null vectors created by the process of adding a floor scales with $\sqrt{N}$, the linear dimension of the system. As the number of beads $N$ in the system increases, these null vectors become a vanishingly small fraction of the number of modes in the system. This scaling is shown in Figure~\ref{Fig:NumberNullVectors}.

The second problem caused by the creation of a null space of $M$ is more worrisome. Because the system is hypostatic at the top, $M$ is rank-deficient. This means that the space of all vectors $M\nkv f$ has a smaller dimension than the space of possible body forces $\nkv F$. Thus there is only an infinitesimal probability that some $\nkv F$ will actually have a contact force network $\nkv f$ which can support it; the system is unstable. In particular, it is in general impossible to find any solution to the problem of supplying a point force. 

\subsection{\label{Subsec:ClusterPull}Applying forces to a local cluster of beads}

\begin{figure}
\begin{tabular}{cc}
\resizebox{40mm}{!}{\includegraphics{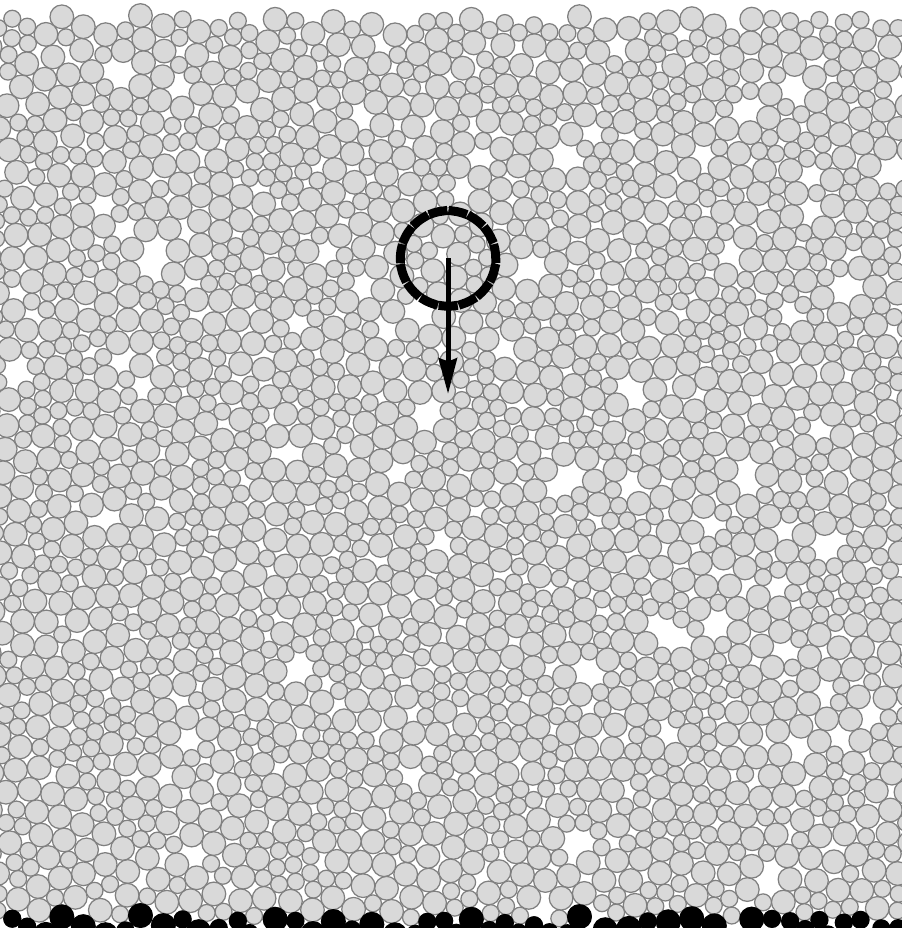}} & 
\resizebox{40mm}{!}{\includegraphics{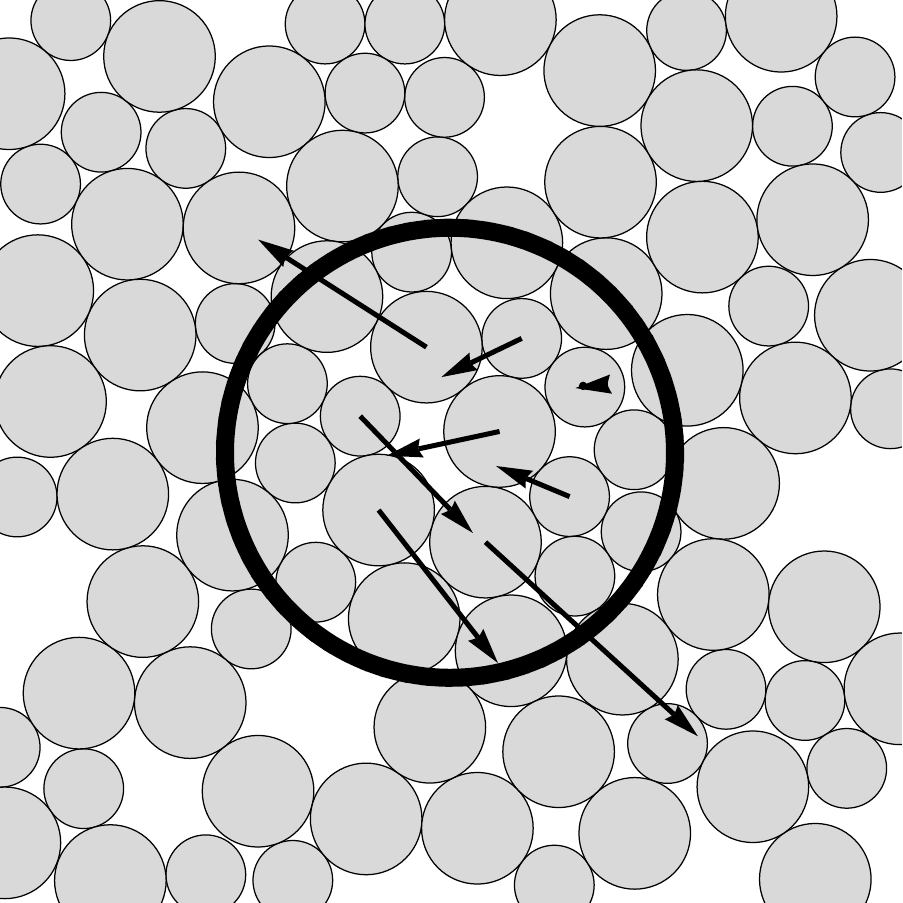}}
\end{tabular}
\caption{\label{Fig:ClusterPull}Instead of supplying a force to a single bead, we can apply forces as shown to a small cluster of beads. In this case, the pack's contact matrix had a 16 dimensional null space, so we applied $x$ and $y$ forces to  8 beads in the circle indicated in on the left, to give the net downward force indicated. On the right, we show the individual forces applied to the beads. Their magnitudes were chosen so as to minimize $|\nkv f|^2$, as discussed in Section~\ref{Sec:Background}.}
\end{figure}

One way to guarantee that the system will be stable under an applied force is to modify the force so that it is in the image of $M$. This can be done by applying external forces to several beads in a small localized cluster, rather than to a single bead. If the null space of $M$ has dimension $k$, then the vectors $M\nkv f$ span a space of dimension $2N-k$ with $N$ the number of beads in the system. Thus, some combination of $k$ linearly independent forces can be added together to give an $\nkv F$ that is in this $2N-k$ dimensional image of $M$. We can take the linearly independent forces to be, for example, vertical and horizontal forces on $k/2$ beads in a small region. Furthermore, if we use $k+2$ independent forces, then we can specify the magnitude and direction of $\nkv F$ as well. This lets us supply a force similar to the point force used in the sequential packs, though it is spread out into a slightly larger area, as shown in Figure~\ref{Fig:ClusterPull}. Since $k$ scales like $\sqrt{N}$, this cluster becomes a better approximation of a point force as the system size is increased. Strangely, it is typically possible to obtain an $\nkv F$ in the image of $M$ with far fewer than $k+2$ independent forces. This must be due to some special feature of $M$'s structure, as this is not true for an arbitrary matrix. We have unfortunately been unable to determine what this feature is. Nevertheless, using this method, the contact forces can be found for these packs, allowing the computation of a well-defined $\nkv G$.

As shown in Figure~\ref{Fig:AveragingResponses}a, for an individual pack the response looks quite wild, and many realizations need to be averaged together for the continuum behavior to be apparent. To get an idea of how many realizations are needed, we can construct a function $c$, defined by
\begin{equation}
\label{Eqn:Noise}
c(N_R,y_0) = \frac{\sum_{x_i}|G_1^{(N_R)}(x_i,y_0)-G_2^{(N_R)}(x_i,y_0)|}{\sum_{x_i}|G_1^{(N_R)}(x_i,y_0)+G_2^{(N_R)}(x_i,y_0)|}
\end{equation}
where $G_1^{(N_R)}$ and $G_2^{(N_R)}$ are each responses averaged over different sets of $N_R$ realizations. For any particular $y_0$, this indicates how closely the two different functions match. If $c(N_R,y_0)$ is small, then both sets of $N_R$ realizations average to something similar, and $N_R$ is large enough to see the continuum behavior. 

\begin{figure}
\resizebox{80mm}{!}{\includegraphics{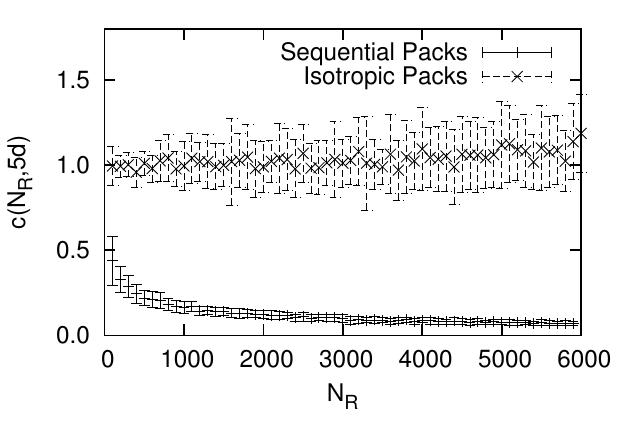}}
\caption{\label{Fig:NoiseFunction}The function $c$ of Equation~\ref{Eqn:Noise} for both sequential and isotropic packs, a distance of five bead-diameters below the location of the applied force. $c$ depends slightly on the particular choice of realizations used to compute $G_1^{(R)}$ and $G_2^{(R)}$. The error bars show the spread of values obtained over 25 different choices. For sequential packs, $c$ decreases as $1/\sqrt{R}$. For isotropic packs, it shows no sign of decreasing, even over three decades.}
\end{figure}

For the sequential packs discussed in Section~\ref{Sec:SequentialPacks}, $c$ scales like $1/\sqrt{N_R}$, as could be expected. However, as seen in Figure~\ref{Fig:NoiseFunction}, $c$ does not decrease with $N_R$ for the isotropic packs, indicating that averaging does not give any convergence; there seems to be no continuum behavior that would allow one to speak of a coarse-grained stress field.

This lack of averaging is possible due to the distribution of contact force sizes. Disorder in isostatic systems causes multiplicative rather than additive noise. Systems with this sort of noise do not average well: the Green's functions have magnitudes that vary in a range that grows exponentially with distance from the applied force, and follow a power law distribution $P(G)\sim G^{-\alpha}$, with $\alpha$ close to one \cite{Moukarzel2002}. Thus in any collection of packs, the magnitudes of the contact forces can vary exponentially, and a small number of packs with large forces end up dominating the average response \cite{Redner1990}. At any fixed depth below the forcing point, Newton's Laws assure us that the total force summed across the entire pack width is equal to the supplied force. However, the higher moments diverge, so any averages over a small portion of the pack, rather than the whole width, can fluctuate wildly, making it impossible to measure a continuum response.

\subsection{\label{Subsec:ModifyBoundaries}Modifying the boundaries}

\begin{figure}
\begin{tabular}{c}
\resizebox{80mm}{!}{\includegraphics{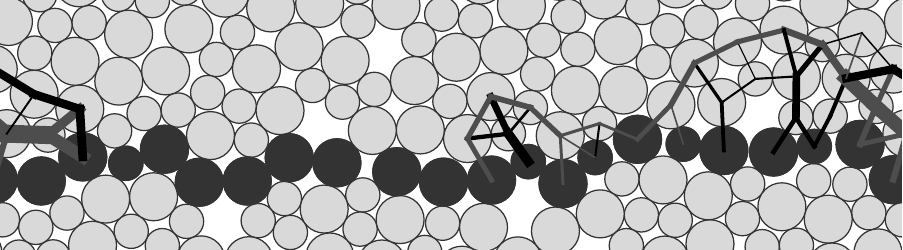}} \\
\resizebox{80mm}{!}{\includegraphics{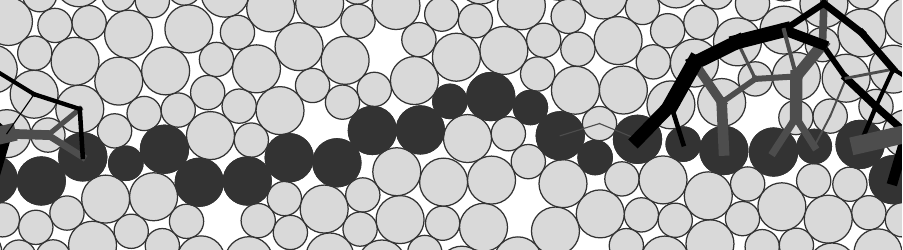}} \\
\resizebox{80mm}{!}{\includegraphics{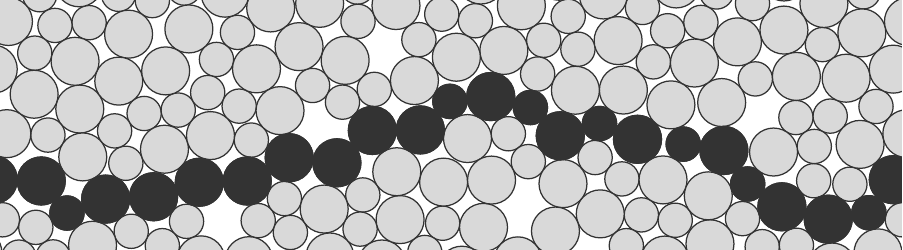}}
\end{tabular}
\caption{\label{Fig:ModifyFloor}On top, a floor made from a straight horizontal line across a pack. The null space for this system is six dimensional. A combination of the null vectors is shown using black and gray lines. Each line represents a contact force between the two beads, whose magnitude is proportional to the line width and whose sign is indicated by the color. The null vector shows a way to have nonzero contact forces, but nevertheless have zero net force on every bead. In the middle image, the center of the floor has been deformed to eliminate part of the null space. The resulting system's null space is only four dimensional. On the bottom, the floor has been further modified so that there is no null space; the contact matrix is now invertible.}
\end{figure}

The null modes of the previous section appear in the process of adding the floor. Instead of using a straight horizontal line to determine the floor beads, more care can be used to create a floor which does not lead to any null modes. To find such a floor, we use a simulated annealing algorithm to minimize the dimension of the null space by making small deformations to a starting boundary, as demonstrated in Figure~\ref{Fig:ModifyFloor}. 

The simulated annealing algorithm starts by creating a directed graph $\mathcal G$ whose nodes are the beads in the pack and whose edges are the contacts, directed from the left to the right. The periodic boundary conditions in the horizontal direction allow any path on this graph from a bead back to itself to be treated as a floor. Making the graph edges point from right to left will also work as long as the choice of direction is consistent; the idea is to create a floor that spans the entire horizontal length of the system instead of immediately looping back on itself.

Given any floor, the size of the null space of $M$ can be computed. This can be used as some sort of effective energy $\mathcal E$ that we are trying to minimize. Neighboring floors are chosen at random by picking two nearby beads $i$ and $j$ in the current floor and finding the shortest path between them on the graph $\mathcal G'$, which is $\mathcal G$ with all edges on the current path between $i$ and $j$ removed. On rare occasions the new path will be much longer than the original. This is typically a sign that the path wraps around the periodic boundaries multiple times, and so such paths are discarded. The new path will have some energy $\mathcal E'$ associated with it. We keep this path with probability
\begin{equation}
P = \left\{\begin{array}{ll}e^{-(\mathcal E'-\mathcal E)/T} & \textrm{ if } \mathcal E' > \mathcal E\\ 1 & \textrm{ otherwise} \end{array}\right.
\label{Eqn:BoundaryProbability}
\end{equation}
The effective temperature is chosen to go like $T\sim1/\log(s)$, with $s$ the number of steps taken so far. This temperature function can probably be improved, but the process finds states where $\mathcal E=0$ fairly rapidly as it is.

When a floor with $\mathcal E=0$ is found, the algorithm will remember it and continue to run until some set number of steps, typically on the order of $10^3$, is reached. It is reasonable to wonder whether the existence of floors which create no null space is dependent on having a finite-sized pack. This turns out not be the case, however. In fact, the larger the pack, the more likely the existence of such a floor. This is shown in Appendix~\ref{Sec:GoodFloors}. Thus, using a variety of starting points, many floors that do not cause a null space can typically be found. As discussed in Section~\ref{Sec:Background}, we expect the lowest modes of $\mathcal D$ to be the most important for determining static force distributions. From the set of possible floors, we choose the one whose lowest eigenvalues have magnitudes similar to those found in sequential packs of the same size.

Even in this case, where the important eigenvalues of the isotropic packs are similar to those of the sequential packs, the response varies too wildly to give an average. One cannot speak of any continuum stress field, let alone one which obeys a null stress condition. 

\section{\label{Sec:HardTop}Stabilizing the isotropic packs}

In the previous section, we showed that periodic isotropic packs can be modified to have a floor. Though this creates some issues by making $M$ singular, these problems can be overcome to give a well-defined force response for a given pack. However, the resulting forces vary so wildly from pack to pack that it is impossible to infer a continuum average. Here, we make a more drastic change to the pack in an attempt to stabilize it and create a continuum response. We do this by adding a hard ceiling to the pack, in addition to the hard floor. Instead of imposing force balance constraints for the boundary beads (such as beads A and B in Figure~\ref{Fig:AddFloor}) we simply allow them to absorb any force. 

\begin{figure}
\resizebox{80mm}{!}{\includegraphics{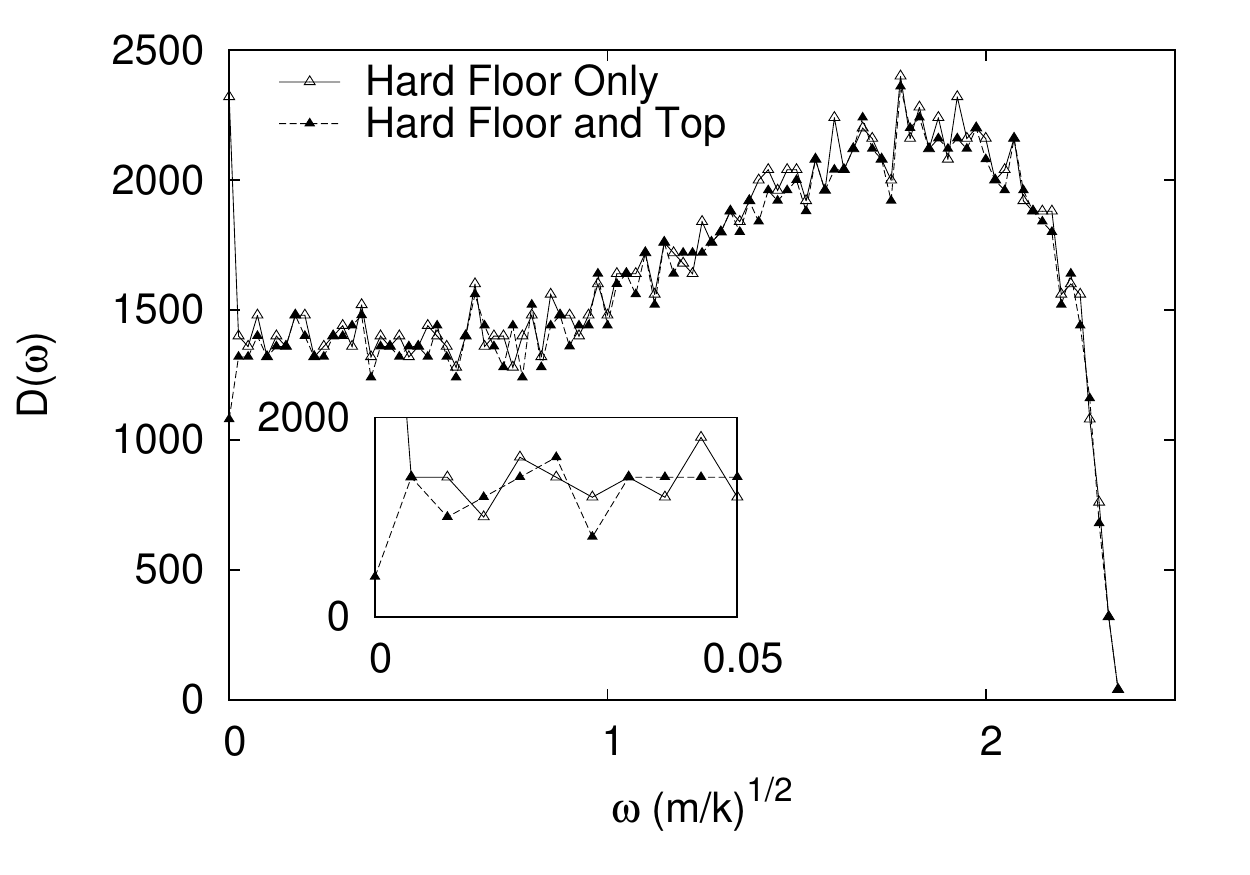}}
\caption{\label{Fig:HardTopDoS}The density of states for a 2000 bead pack before and after a hard ceiling has been added using the method described in Section~\ref{Sec:HardTop}. The general shape is unaffected, and the low frequency plateau expected in an isostatic pack is evident even after the change. The only significant difference is in the first data point, which counts the number of modes below some small $\Delta\omega$. This difference is due to the null modes discussed in Section~\ref{Sec:IsotropicPacks}, which vanish when the hard ceiling is added. As noted in the text, the fraction of modes affected this way becomes vanishingly small as the system size increases.}
\end{figure}

This system is no longer isostatic, as it has excess contacts near both the floor and the ceiling. However, the bulk of the pack is unaffected, and remains isostatic. Adding the ceiling does not greatly affect the density of states; aside from removing the zero eigenvalues, the overall shape of the plateau is not altered. This can be seen in Figure~\ref{Fig:HardTopDoS}. As noted in the previous section, the number of zero eigenvalues of the pack without the hard ceiling goes like $\sqrt{N}$, so the number of affected modes is an asymptotically small fraction of the total number.

\begin{figure}
\resizebox{80mm}{!}{\includegraphics{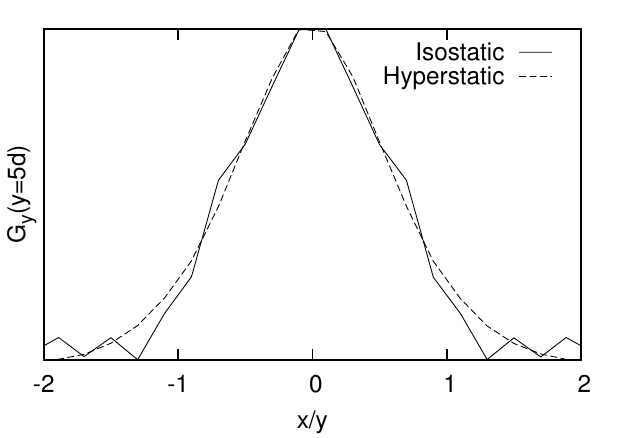}}
\caption{\label{Fig:HardTopResponse}A comparison of the response five bead-diameters below the forcing point for 5584 isostatic packs with the same number of hyperstatic ($Z\sim 5.6$) packs, when hard ceilings are added. The hyperstatic packs were formed by adding contacts to the isostatic ones between all pairs of nearest-neighbor beads. The responses are in arbitrary units, normalized to the maximum value. Both pack types have the same functional form. In both cases, the response more closely resembles an elastic solid, with a single peak below the forcing point, than it does the sequential packs.}
\end{figure}

Unlike the case with a free top, the responses now average nicely. However, instead of forming light cones, the stress now peaks directly below the applied force. This is shown in Figure~\ref{Fig:HardTopResponse}. The response of our isostatic pack with slightly hyperstatic boundaries now resembles that of a system which is hyperstatic everywhere. That is, the average force response behaves more like it would in an elastic solid \cite{Goldenberg2002,Reydellet2001}. The isotropic packs must then in some sense be stronger than the sequential ones, since one can restore elastic-like behavior to them with much less modification.

\begin{figure}
\begin{tabular}{cc}
\resizebox{40mm}{!}{\includegraphics{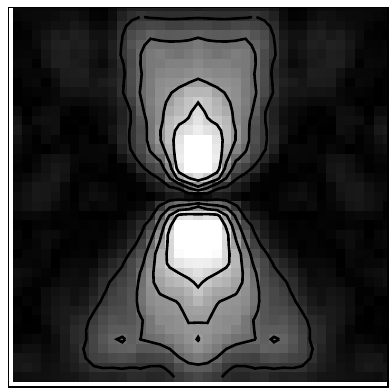}} & 
\resizebox{40mm}{!}{\includegraphics{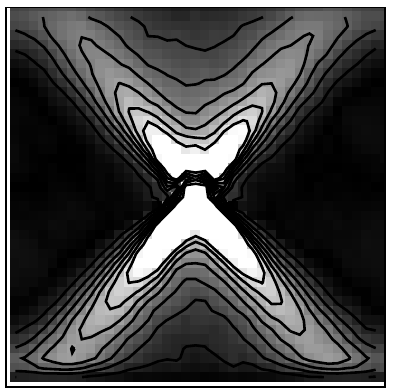}}
\end{tabular}
\caption{\label{Fig:HardTopContours}On the left, a plot of the stress $\sigma_{yy}$ in isotropic packs with hard ceilings, obtained by averaging over 15254 realizations. For comparison, on the right is $\sigma_{yy}$ for the sequential packs when hard ceilings have been added. In both cases, the forcing point is in the center, and is directed down. The light cones are still clearly visible for the sequential packs, indicating that the presence of the ceiling does not cause the elastic-like behavior seen in the isotropic packs.}
\end{figure}

It may be that this new behavior in the bulk is due to the hyperstaticity at the boundaries. However, this should not be the case. Even when the average number of contacts per bead in some pack exceeds the isostatic number by some amount $\Delta z$, it should only behave elastically on length scales larger than $\ell^*\sim 1/\Delta z$ \cite{Wyart2005,Silbert2005}. There is evidence that an ensemble average of many such packs with a small $\Delta z$ will nevertheless show elastic-like behavior on smaller length scales when the contacts are added randomly throughout the system \cite{Ellenbroek2009}. However, this may depend on the ensemble average being able to distribute the resulting elastic energy stored in the extra bonds uniformly throughout the pack. In our system, the excess contacts are all near the boundaries. This means that not only does the fraction of extra contacts $\Delta z\rightarrow 0$ for large packs, but also that any sort of elastic behavior should be limited to the contacts there. Unless the coarse-graining is done on length scales larger than the distance between the ceiling and the floor, there is no way for the extra contacts to affect the behavior in the bulk. We can test this by giving sequential packs the same floors and ceilings as these isotropic ones, which yields similar departures from isostaticity. This does not create a stress profile with a single peak. Instead, the light cones are visible going to both the floor and the ceiling, as shown in Figure~\ref{Fig:HardTopContours}. This further indicates that it is the isostatic bulk of the isotropic packs which cause the elastic-like response, and not the boundaries. 

The lack of light cones in the isotropic packs with hard ceilings is an indication that the null stress law expressed in Equation~\ref{Eqn:NullStressLaw} is not satisfied. Previous verifications of the null stress law have been done by applying a uniform load across the top of the entire pack, and then measuring the average $\sigma_{k\ell}$. By varying the direction of the load, the $\sigma_{k\ell}$ can take on a range of values. Head et. al.~\cite{Head2001} found that $\sigma_{xx}/\sigma_{yy}$ was essentially constant as long as $\sigma_{xy}/\sigma_{yy}\lesssim 0.3$. This was consistent with the null stress law of Equation~\ref{Eqn:NullStressLaw} with $\mu=0$, though it broke down when the shear became large enough that the left-right symmetry in their systems was broken. However, this approach is not sufficient to distinguish null stress behavior from elastic-like behavior. 

To see this, say that the load applied evenly across the top of the $L\times L$ pack is $\nkv F=(F_x,F_y)$. Because there can be no $x$ dependence in this case, Equation~\ref{Eqn:Continuity} reduces to $\partial_y\sigma_{xy}=\partial_y\sigma_{yy}=0$, implying that those components of $\sigma$ also have no $y$ dependence. Their values can be found at the top: $\sigma_{xy}=F_x/L$ and $\sigma_{yy}=F_y/L$. If the null stress condition holds with $\mu=0$, then as $\nkv F$ varies, $\sigma_{xy}/\sigma_{yy}$ changes, but $\sigma_{xx}/\sigma_{yy}=\eta$ remains constant. This is indistinguishable from what happens in a two-dimensional elastic solid. There, the continuity equations give the same values for $\sigma_{xy}$ and $\sigma_{yy}$. To find $\sigma_{xx}$, note that the strain satisfies
\begin{equation}
\label{Eqn:StrainStress}
u_{xx} = \partial_xu_x = \frac{K(\sigma_{xx}-\sigma_{yy})+G(\sigma_{xx}+\sigma_{yy})}{4GK}
\end{equation}
where $K$ and $G$ are the bulk and shear moduli, respectively. The bulk modulus measures the system's resistance to uniform compression, and the shear modulus is the ratio of shear stress to shear strain at constant volume \cite{LandauTheoryOfElasticity}. Since there is no $x$ dependence, Equation~\ref{Eqn:StrainStress} tells us that $\sigma_{xx}/\sigma_{yy}=\nu$, where $\nu=(K-G)/(K+G)$ is the Poisson ratio. Again, as $\nkv F$ varies, $\sigma_{xy}/\sigma_{yy}$ will change, but $\sigma_{xx}/\sigma_{yy}$ will remain constant.

\begin{figure}
\resizebox{80mm}{!}{\includegraphics{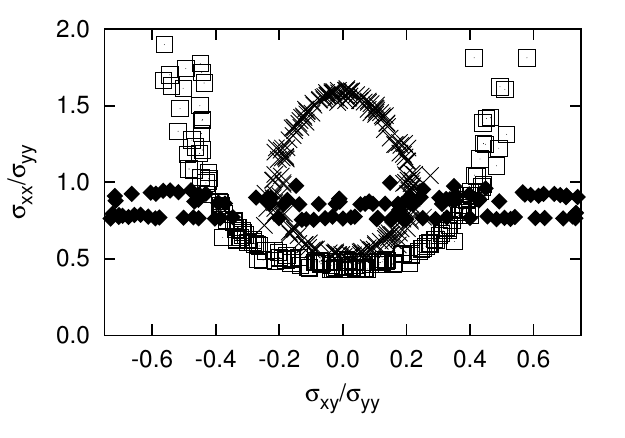}}
\caption{\label{Fig:NullStress}The relationship between the different components of the stress tensor, computed using Equation~\ref{Eqn:StressFromContactForces}, with a coarse graining function of width $w=4d$. Each data point gives the ratio of the stresses at a random location. The black diamonds were obtained by averaging over 12474 realizations of the sequential packs of Section~\ref{Sec:SequentialPacks}, and the squares show the results for isotropic packs with added hard ceilings, as described in Section~\ref{Sec:HardTop}. All packs were made using $N=500$ beads. For comparison, the crosses show the results for an elastic medium, computed using the formulas in Appendix~\ref{Sec:ElasticTheory} The measurements fall in a narrow annulus because the computed values are the average over an area. The elastic medium was given a Poisson ratio of $\nu = 0.9$, which is the value estimated for the isotropic packs of this size (see Figure~\ref{Fig:ModulusDecay}) using the methods described in Section~\ref{Sec:HardTop}. The sequential packs approximately satisfy the null stress condition of Equation~\ref{Eqn:NullStressLaw}, though $\sigma_{xx}/\sigma_{yy}$ evidently varies by up to 25\%; the larger values come from points nearer to the source. The isotropic packs, on the other hand, clearly do not. Though there are some quantitative differences with the elastic media, the general form is the same. While not evident from this figure, the isotropic packs have two distinct values of $\sigma_{xx}/\sigma_{yy}$ where $\sigma_{xy}/\sigma_{yy}$ is zero, just as in the elastic case: one near $\sigma_{xx}/\sigma_{yy} = 0.4$, and another around $\sigma_{xx}/\sigma_{yy}\approx 10$. This latter ratio occurs at points far from the applied force, where $\sigma_{yy}$ is close to zero. Small numerical fluctuations in $\sigma_{yy}$ thus cause large deviations in the computed ratios, so such points are not plotted in the figure.}
\end{figure}

To distinguish between the two situations, we do not supply a load across the entire width of the pack. Instead, we supply a point force as elsewhere in this paper, and then use Equations~\ref{Eqn:StressFromContactForces} and \ref{Eqn:CoarseGrainDefinition} with $w=4$ bead-diameters to find the average stresses at several random locations. For each such area, we then compare $\sigma_{xx}/\sigma_{yy}$ with $\sigma_{xy}/\sigma_{yy}$. If the null stress condition holds, then $\sigma_{xx}/\sigma_{yy}$ should remain essentially constant, no matter the value of $\sigma_{xy}$. Figure~\ref{Fig:NullStress} shows the results for both isotropic packs with hard ceilings and sequential packs. For comparison, we also show the stresses from a periodic two-dimensional elastic solid with hard floors and ceilings. The elastic stresses were computed using the results of Leonforte, et. al. \cite{Leonforte2004}, who extended the work of Serero et. al. \cite{Serero2001}. The relevant formulas are quoted, using our notation, in Appendix~\ref{Sec:ElasticTheory}. We find that in the isotropic packs with added hard ceilings, there is clearly no linear relationship between the components of the stress tensor. The sequential packs, on the other hand, seem to be more in line with the null-stress behavior we anticipated. However, this test is only intended to demonstrate the clear contrast between the sequential packs and the isotropic ones; we have not unambiguously identified the source of this contrast.

\subsection{\label{Sec:EffectiveModuli}Effective moduli of isotropic packs}

Since the null stress relationship between the components of the stress tensor is not satisfied for the isostatic packs with hard ceilings, we can ask what other constitutive equation might describe them. As noted above, the response in many ways resembles what one would see in an elastic solid. Thus, we can compare the results to linear elasticity and attempt to find effective moduli.

To compute the moduli, we supply a vertical force $\nkv F = F\nkh y$ to each of the beads that are intersected by a horizontal line through the middle of the pack. This applied pressure will cause a discontinuity in the $yy$ stress. Above the line it will take some value $\sigma_{yy}$, and below the line it will be $-\sigma_{yy}$. In a two dimensional elastic solid, the stress-strain relationship is
\begin{equation}
\label{Eqn:StressStrain}
\sigma_{\alpha\beta} = K u_{\gamma\gamma}\delta_{\alpha\beta} + 2 G\Big(u_{\alpha\beta}-\frac 12 u_{\gamma\gamma}\delta_{\alpha\beta}\Big)
\end{equation}
where repeated indices are summed \cite{LandauTheoryOfElasticity}. The factor of $1/2$ comes from working in two dimensions rather than three. We do not apply any shear with our force, so $\sigma_{xy}$, and thus $u_{xy}$, are both zero. Furthermore, the applied force does not allow any $x$ dependence, so $u_{xx}=\partial_xu_x=0$. This leaves us with
\begin{align}
\sigma_{xx}&=Ku_{yy}-Gu_{yy}\\ \notag
\sigma_{yy}&=Ku_{yy}+Gu_{yy} 
\end{align}
Dividing gives
\begin{equation}
\label{Eqn:ComputePoissonRatio}
\frac{\sigma_{xx}}{\sigma_{yy}} = \frac{K-G}{K+G}
\end{equation}
Thus $\sigma_{xx}/\sigma_{yy}$ is the Poisson ratio, just as it was in the case of the unconstrained top discussed earlier.

To determine the bulk and shear moduli separately, we measure the energy stored in the pack. The energy density in a two-dimensional elastic solid can be written in terms of the strains as \cite{LandauTheoryOfElasticity}
\begin{equation}
\label{Eqn:ElasticEnergyDensity}
e = \frac 12 K u_{\gamma\gamma}^2 + G\Big(u_{\alpha\beta}-\frac 12 u_{\gamma\gamma}\delta_{\alpha\beta}\Big)^2
\end{equation}
with the factor of $1/2$ again coming from the dimensionality of the system. From above, $u_{xx}=u_{xy}=0$ and $u_{yy}=\sigma_{yy}/(K+G)$. The energy stored in the entire pack is therefore
\begin{equation}
\mathcal E = L^2e = \frac {L^2}{2}\frac{\sigma_{yy}^2}{K+G}
\end{equation}
We can also use $\mathcal E = \frac{1}{2k}\nkv f^T\nkv f$ from Equation~\ref{Eqn:ConvertMtoD} to compute the energy stored in response to our force. Equating the two gives
\begin{equation}
K+G = k\frac{L^2 \sigma_{yy}^2}{|\nkv f|^2}
\end{equation}
Combining this with Equation~\ref{Eqn:ComputePoissonRatio} lets us solve individually for the moduli:
\begin{align}
\label{Eqn:IndividualModuli}
K &= \frac{k}{2}\frac{L^2\sigma_{yy}^2}{|\nkv f|^2}\Big(1+\frac{\sigma_{xx}}{\sigma_{yy}}\Big) \\ \notag
G &= \frac{k}{2}\frac{L^2\sigma_{yy}^2}{|\nkv f|^2}\Big(1-\frac{\sigma_{xx}}{\sigma_{yy}}\Big)
\end{align}
Note that as $K$ gets large compared to $G$, the formula for $G$ becomes less accurate. An alternative form of forcing, $\nkv F = F\nkh x$, would provide a shear which is not dominated by the effects of the bulk modulus.

Since $\sigma_{xx}$ and $\sigma_{yy}$ are constant on either side of the applied force, they can be computed using Equation~\ref{Eqn:StressFromContactForces} with a $\Phi$ that encompasses either the top or bottom half of the pack. We expect the area away from the boundaries and the applied force to be better-behaved, so we can imagine an $L\times 3L/10$ area $A$ which includes the half of the pack between the applied force and the ceiling, excluding anything within $L/10$ of either of those boundaries. Then
\[
\sigma_{\alpha\beta} = \sum_{(i,j)} f_{ij}R_{ij}(n_{ij})_\alpha(n_{ij})_\beta\Bigg[\frac{1}{L(\frac{3L}{10})}\left\{\begin{array}{ll}1;&(i,j)\in A\\0;&\textrm{else}\end{array}\right.\Bigg]
\]

A perfectly ordered isostatic lattice can have finite moduli, but in disordered isostatic systems, they approach zero as the system size increases. Moukarzel \cite{Moukarzel2012} has presented numerical data suggesting that the form of the moduli depends on the packing symmetries. For isotropic isostatic networks with periodic boundary conditions in all directions, he found that they decrease like $A N^{-B}$, where $B$ depends on the pack construction and disorder. However, for packs constructed from a hard floor with a preferred direction, the decay is exponential. He argues that this is caused by the multiplicative noise discussed in Section~\ref{Subsec:ClusterPull}.

\begin{figure}
\resizebox{80mm}{!}{\includegraphics{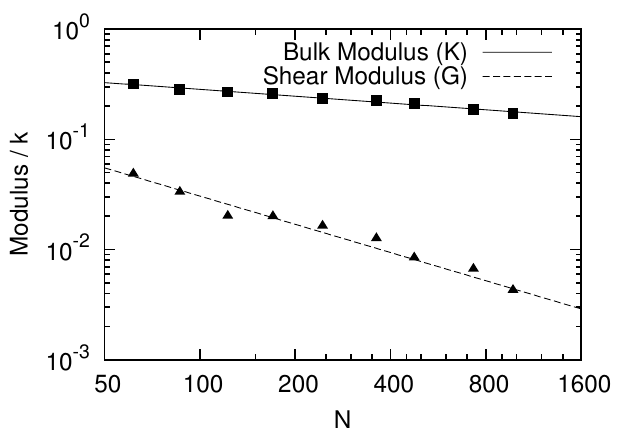}}
\caption{\label{Fig:ModulusDecay}A log-log plot of the bulk and shear moduli of the isotropic packs with hard ceilings and floors of Section~\ref{Sec:HardTop} as a function of the number of beads $N$ which make up the pack. The moduli are normalized by the spring constant $k$ of the harmonic force between contacting beads. Each data point is the average of at least 100 different realizations of the same size. The dashed line is a fit to the form $AN^{-B}$, with $B_K=0.21\pm 0.1$ for the bulk modulus, and $B_G=0.85\pm0.07$ for the shear modulus. The beads in the packs have a bidispersity of 1.4 : 1. Moderate changes to the bidispersity do not affect the scaling.}
\end{figure}

The systems we consider here share the isotropic bulk of the power-law systems, and the preferred boundary directions of the exponential systems. Figure~\ref{Fig:ModulusDecay} shows that moduli follow the power-law behavior of the isotropic packs, with exponents of $B_K=0.21\pm0.01$ for the bulk modulus and $B_G=0.85\pm0.07$ for the shear modulus. With the added hard ceiling, our system does not display the exponentially large contact forces, supporting the claim that it is the multiplicative noise and not the boundary conditions which cause an exponential decay. An alternative explanation given for this elastic collapse in Moukarzel's systems with average contact number of exactly $Z=4$ is that such systems are not actually jammed, due to an extra degree of freedom which comes from the system size \cite{Goodrich2012}. Since the systems are not jammed, vanishing moduli are expected. However, our systems are over-constrained by more than one contact, so the single degree of freedom used in this argument does not seem to explain our decay, though it is possible that we have simply not gone to large enough $N$.

Because the shear modulus decays faster than the bulk modulus, the Poisson ratio approaches 1 in the limit of infinite system size. This is the expected value of $\nu$: in disordered packs with small excess contact number $\Delta z$, the bulk modulus is proportional to the effective spring constant $k$, whereas the shear modulus is proportional to $k\Delta z$ \cite{Ellenbroek2009}. In two dimensions then, $\nu=(K-G)/(K+G)\rightarrow 1$ as $\Delta z\rightarrow 0$.

These systems with added hard ceilings show that it is possible to have a continuum response in an isostatic system with no light cones. The stress can instead follow a pattern that can be described using linear elasticity, with bulk and shear moduli which exhibit power law decays when the system size increases. The presence of elastic-like behavior indicates that light cones do not arise from simple isostaticity. More specific geometric concerns must also play a role. These are discussed in the next section.

\section{\label{Sec:ContactAngles}The Contact Angle Distribution}

Though many properties of sequential and isotropic packs are similar, their contact angle distributions are quite different. For a pair of beads, we can look at the angle $\theta$ which the line between their centers forms with respect to a vertical line from the floor. By doing this for every pair of contacting beads in many packs, we get a probability distribution $P(\theta)$.

\begin{figure}
\begin{tabular}{cc}
\resizebox{30mm}{!}{\includegraphics{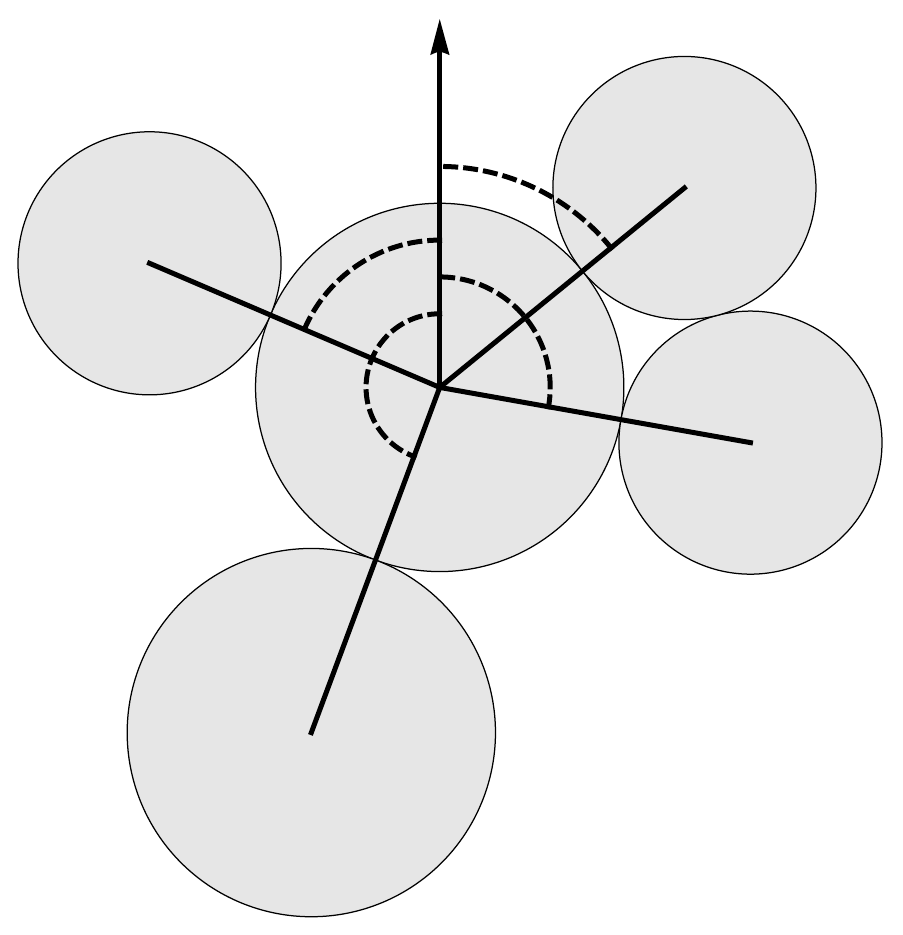}} &
\resizebox{50mm}{!}{\includegraphics{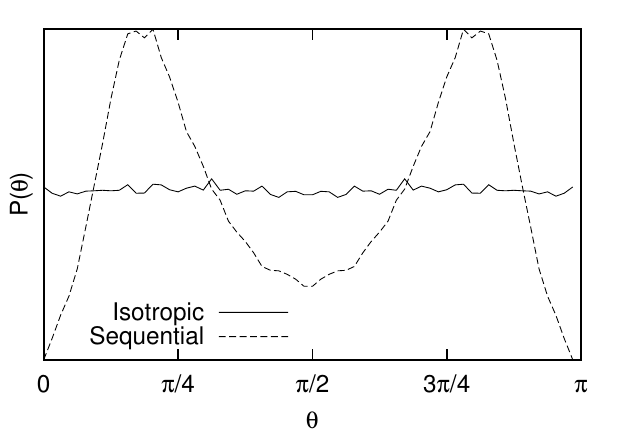}}
\end{tabular}
\caption{\label{Fig:PeakedPofTheta}On the left, we show the contact angles for a bead. They are defined to be the angle between each contact and a line pointing up from the floor, as shown by the dashed lines. Note that if for some bead we measure an angle $\theta$, then for the contacting bead, we will measure $\pi-\theta$. The distribution is therefore symmetric about $\pi/2$. On the right, we see the probability density function $P(\theta)$ found by computing all $\theta$ values in 100 packs, for both the sequential and isotropic cases. While the isotropic packs have no preferred angle, the sequential packs have strong peaks. Due to a preference for horizontal rather than vertical contacts, the average contact angle is slightly to the side of these peaks, close to $\langle\theta\rangle\sim 45^\circ$. }
\end{figure}

In our isotropic packs, $P(\theta)$ is a constant. However, in the sequential packs, there are peaks near 45$^\circ$ and 135$^\circ$, as shown in Figure~\ref{Fig:PeakedPofTheta}. This is the same angle at which the light cones propagate. The correlation of light cone direction with bond angle suggests a more prosaic explanation for the light cones than the formal null-stress picture. Light cones may arise in sequential packs simply because each bead is sitting on top of two others, which on average will be in the directions the light cones go.

We can attempt to test this by varying the distribution of the bead sizes in sequential packs. This will modify the probability density function $P(\theta)$ of the contact angle distribution, which can be compared to the positions of the light cones. In addition to a few polydisperse systems where the radius of each bead is chosen from the flat distribution in the range $[R,\alpha R]$, we also use several bidisperse systems where the radius is either $R$ or $\alpha R$. In the bidisperse cases, half of the beads are given each radius, so that the larger beads occupy more total area.

\begin{figure}
\resizebox{80mm}{!}{\includegraphics{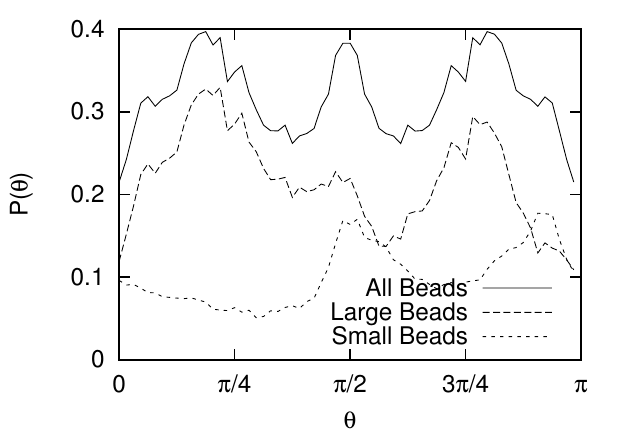}}
\caption{\label{Fig:VeryBidisperseAngles}The probability density function (PDF) of contact angles $P(\theta)$ for a pack with a large bidispersity. The radii of the large beads are 3.5 times bigger than those of the small ones. The solid line shows the PDF using all of the contacts in 100 packs, and is symmetric about $\pi/2$. The two other curves show the PDF for only the contacts of beads of a particular size, weighted by the fraction of the total number of contacts in the pack with beads of that size. These two curves sum to the PDF for the entire pack, but are not individually symmetric.}
\end{figure}

For the more extreme bidispersities, $P(\theta)$ no longer has the form of two clean peaks, as shown in Figure~\ref{Fig:VeryBidisperseAngles}. However, most of the deviation from that form is due to the small beads, which end up not being important for the force transmission. Many of them serve only to fill in the spaces between large beads, and do not experience any contact force; while typically about 80\% of the large beads in a pack experience some nonzero contact force, only about 15\% of the small beads do. Thus when computing the average contact angle for comparison with the light cone direction, we use the PDF for the downward propagation from the large beads only, since they are the ones dominating the contact force network. 

Table~\ref{Table:AnglesVsPeakPositions} compares the average $\theta$ in the range $[0,\pi/2]$ with the angle $\varphi$ the light cones make with the vertical, for a range of radius distributions. While $\theta$ is consistent with $\varphi$ for the cases shown, the range of possible angle distributions that can be obtained from sequential packs in this way is quite limited, and any shifts in the peak positions are within uncertainties.

\begin{table}
\begin{tabular}{|l|d|d|d|}
\hline
Type & \alpha & \langle\theta\rangle & \varphi \\
\hline
\multirow{5}{*}{Bidisperse} 
	& 1.4 & 43.4\pm 0.6^\circ & 40\pm 3^\circ\\
	& 2.0 & 41.7\pm 0.6^\circ & 38\pm 4^\circ\\
	& 2.5 & 39.4\pm 1.0^\circ & 40\pm 5^\circ \\
	& 3.0 & 42.8\pm 1.1^\circ & 40\pm 5^\circ \\
	& 3.5 & 45.8\pm 1.1^\circ & 38\pm 4^\circ \\
\hline
\multirow{3}{*}{Polydisperse} 
	& 1.1 & 44.0\pm 0.4^\circ & 44\pm 2^\circ\\
	& 2.0 & 43.0\pm 0.4^\circ & 40\pm 3^\circ\\
	& 3.0 & 42.5\pm 0.6^\circ & 40\pm 5^\circ\\
	& 6.0 & 43.5\pm 0.6^\circ & 39\pm 4^\circ\\
\hline
\end{tabular}
\caption{\label{Table:AnglesVsPeakPositions}A comparison of the light cone positions $\varphi$ with the average contact contact angle $\langle\theta\rangle$, for several different pack types. All of the packs used here were created using the sequential method of Section~\ref{Sec:SequentialPacks}. For the bidisperse packs, half of the beads were given radius $R$, and the other half had radius $\alpha R$. For the polydisperse packs, the radii were chosen randomly from the uniform distribution $[R,\alpha R]$. In the bidisperse packs, the disorder in the pack initially increases with $\alpha$, creating wider and less defined light cones. As $\alpha$ increases further, the small beads become insignificant compared to the larger beads, and the system becomes more ordered; the small beads play only a minimal role in both the overall structure of the pack and the force transmission. This is discussed in Figure~\ref{Fig:VeryBidisperseAngles}. The polydisperse peaks tend to be more clearly defined even at large polydispersities, as there is no sharp separation of beads based on size.}
\end{table}

\begin{figure}
\begin{tabular}{ccc}
\resizebox{27mm}{!}{\includegraphics{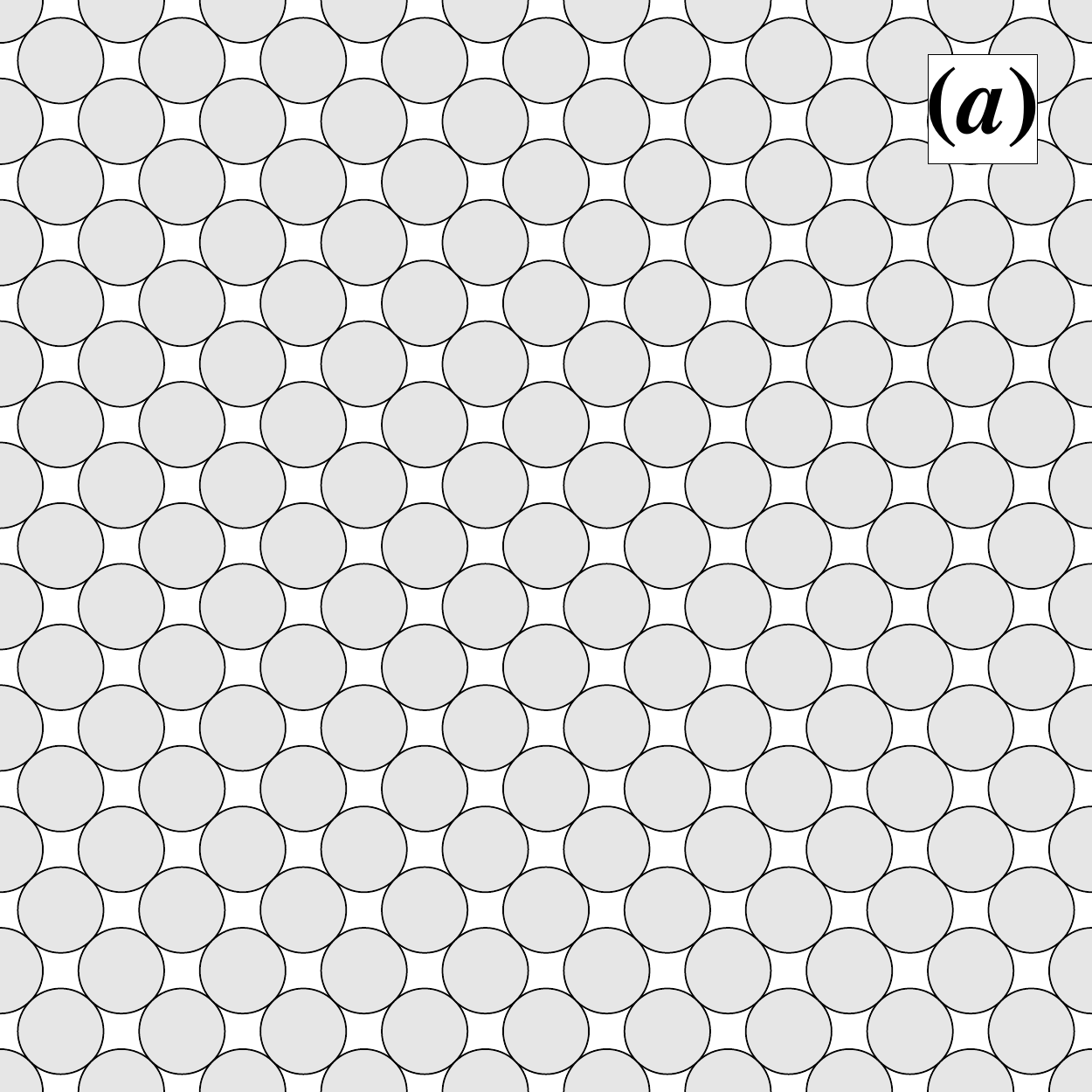}} & 
\resizebox{27mm}{!}{\includegraphics{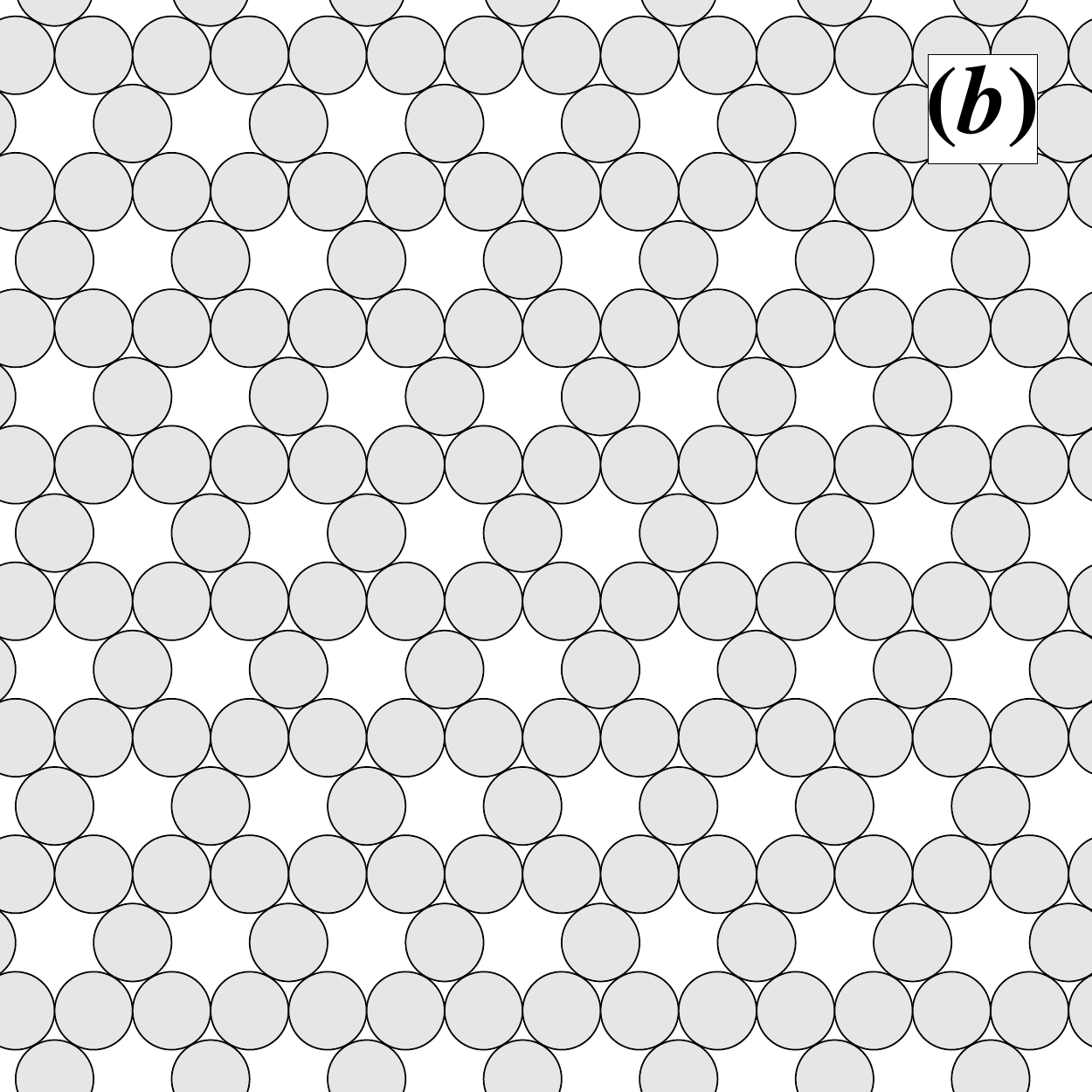}} &
\resizebox{27mm}{!}{\includegraphics{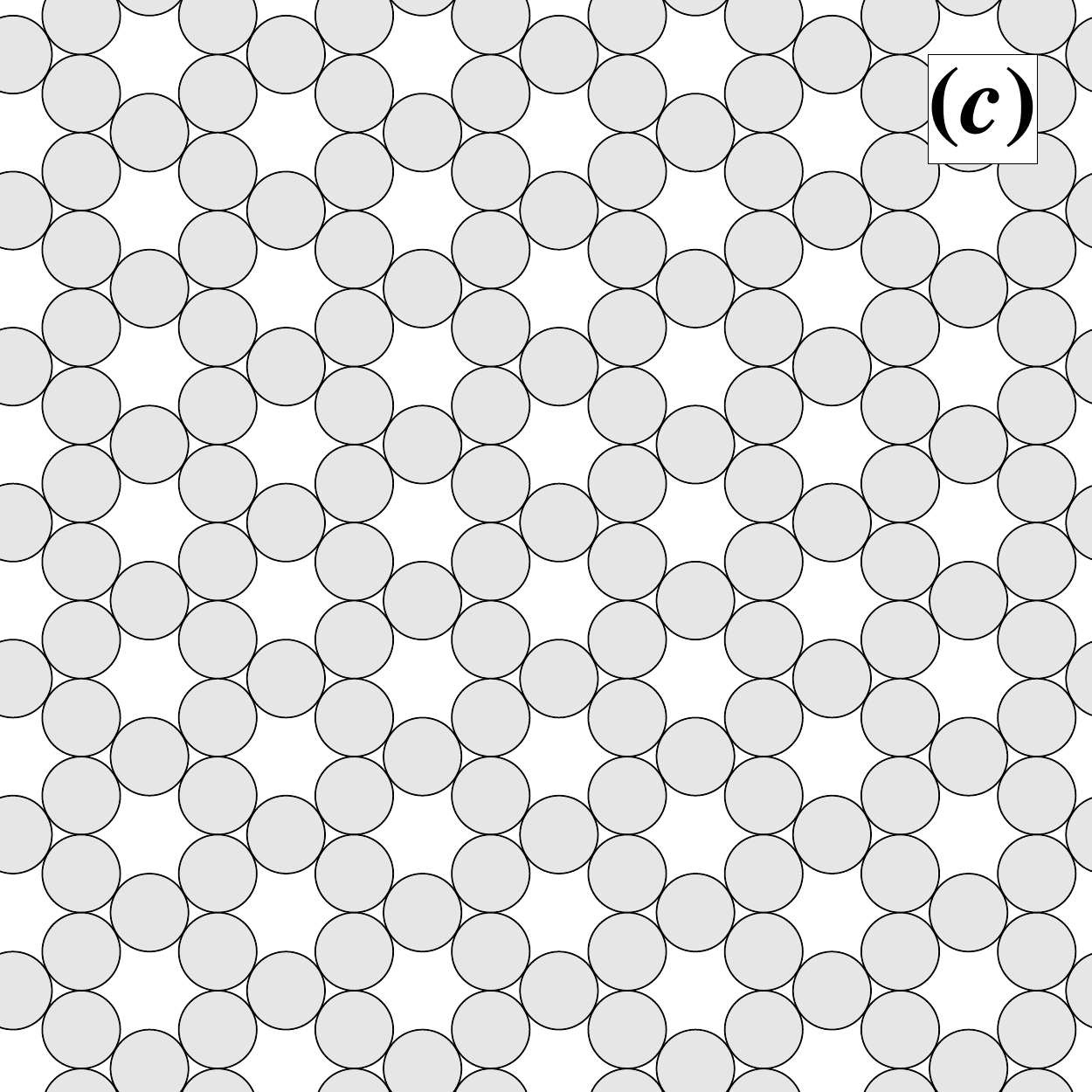}} 
\end{tabular}
\caption{\label{Fig:IsostaticLattices}A diagonal lattice (a) and two different orientations of a kagome lattice (b and c). All of these arrangements have the isostatic contact number. When they are used, randomness is added in the form of modifications to their beads' positions. Bead displacements in the $x$ and $y$ directions are selected at random from a normal distribution of width $\beta R$ for $R$ the bead radius and $\beta$ some positive number. This randomness serves not only to spread out the contact angle distribution, but also to remove the null modes inherent in the unperturbed kagome lattices.}
\end{figure}

To generate more extreme changes in the contact angle distribution, we look at packs formed from isostatic lattices. For each of the base lattice configurations shown in Figure~\ref{Fig:IsostaticLattices}, we perturb the position of each bead by moving both its $x$ and $y$ coordinates a random amount chosen from a normal distribution of width $\beta R$. This causes some beads to overlap, but we ignore this and use only the contact angles to construct $M$ and find the response. This effectively treats the system as point particles connected by springs, and the positional randomness serves merely to modify the rest lengths and angles of these springs.

\begin{figure}
\begin{tabular}{cc}
\resizebox{41mm}{!}{\includegraphics{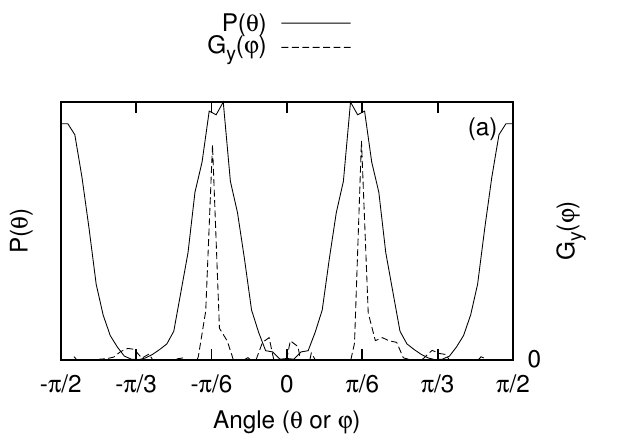}} & 
\resizebox{41mm}{!}{\includegraphics{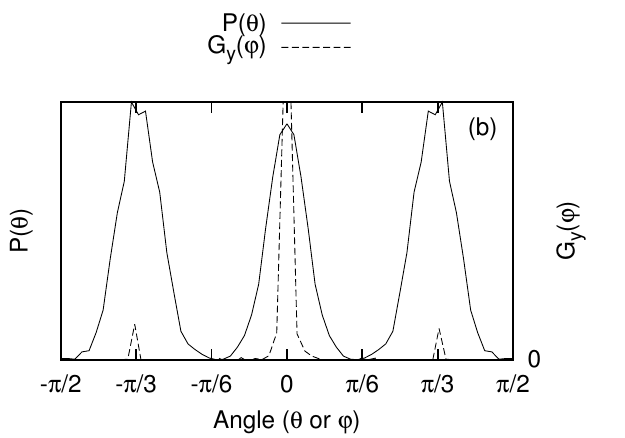}}
\end{tabular}
\caption{\label{Fig:KagomeAngles}A comparison of the contact angle distribution with the light cone direction for the two different orientations of the kagome lattice. In both cases the positions of all beads have been perturbed by an amount chosen from a normal distribution with a width of one fifth of a bead radius. (a) shows the results for the orientation depicted in Figure~\ref{Fig:IsostaticLattices}(b). The two non-horizontal peaks in $P(\theta)$ are at $\langle\theta\rangle\sim\pm 30^\circ$, and the light cones clearly propagate at the same angle. (b) shows the results for the orientation depicted in Figure~\ref{Fig:IsostaticLattices}(c). There is a peak in $P(\theta)$ at $0/\pi$, which gives the strong peak in the center. The other peaks in $P(\theta)$ are at $\langle\theta\rangle\sim 60^\circ$, which agrees with the angles of the smaller light cones. The middle peak is larger because forcing points were chosen at random, and most of the possible choices were on one of the vertical struts. Both sets of response data were obtained by averaging over around 450 realizations of $N=2340$ beads.}
\end{figure}

Floors are added to these packs as shown in Figure~\ref{Fig:AddFloor} for the isotropic case. The diagonal lattice is not so interesting; the average contact angle is still 45$^\circ$, and this is also the angle of the light cones. However, the kagome lattices do show different distributions, which are depicted in Figure~\ref{Fig:KagomeAngles}. In the orientation of Figure~\ref{Fig:IsostaticLattices}b, there are three peaks in the contact angle distribution. One corresponds to horizontal contacts, and the others are around $\langle\theta\rangle=30.0\pm 1.4^\circ$. The light cones in these kagome packs are located at $\varphi=30\pm 3^\circ$.
In the other orientation, one of the peaks in the contact angle distribution corresponds to vertical contacts. The origin of this is obvious from Figure~\ref{Fig:IsostaticLattices}c. For each realization used in the average, the applied point force was exerted on a random bead. Thus a large proportion of the forcing points were located on one of these vertical struts, since that is where most of the beads are. In this case the simplest direction for the force to go is straight down the strut. This gives the large peak in the response corresponding to light cones at $\varphi=0$. The other light cones are at $\varphi=60\pm 3^\circ$. These correspond to the average value of the other peaks in $P(\theta)$, at $\langle\theta\rangle=59.8\pm 2.9^\circ$.

Thus in the isostatic lattices we also find that the light cone angle follows the average contact angle. Given the ordered nature of lattices, this is not as surprising as it was in the much more disordered sequential packs. Experimental studies have shown that the forces in regular granular lattices such as hcp and fcc crystals indeed move in rays following the preferred lattice directions, without any need for isostaticity \cite{Mueggenburg2002}. When a small amount of disorder was added to such systems, there was a transition to a broad central peak after a few layers. Here, we see the light cone-like behavior at all depths in our isostatic lattices, even with significant disorder. Admittedly, our packs are small enough that we can only verify this to a depth of about 20 bead diameters. However, since we also see the same light-cone like behavior in the disordered sequential packs which have no obvious underlying lattice structure, we believe that in this case, it is the isostatic properties rather than the lattice ones which cause the light cones. Nevertheless, it would be nice to have an example of a system with both a substantially different $\langle\theta\rangle$ from what is seen in sequential packs, and no clear lattice structure.

\begin{figure}
\resizebox{80mm}{!}{\includegraphics{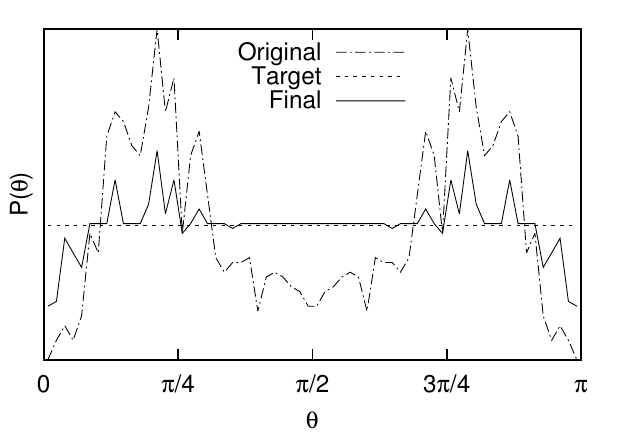}}
\caption{\label{Fig:FixContactAngles}The original contact angle distribution for a single sequential pack, along with the target (flat) distribution that we want to achieve by changing contacts according to the algorithm described in Section~\ref{Sec:ContactAngles}. The final distribution after the angle modification algorithm has completed still shows traces of a peak, but is on a whole much flatter.}
\end{figure}

We attempt to find such systems by taking the disordered sequential packs and modifying their contact networks to give some other target $P(\theta)$. We do this by constructing a list of the nearest neighbors of each bead. We then try to randomly remove one existing contact from the pack, and then add another one from the list of nearest neighbor pairs that are not in contact. If this move makes the contact matrix $M$ singular or takes our $P(\theta)$ further from the target distribution, then we reject it and try again. Otherwise, we keep that move and make another one. This can be continually repeated. In practice, we continue attempting these contact swapping operations until less than one percent of them are accepted. This lets us get very close to the target distribution, as shown in Figure~\ref{Fig:FixContactAngles}.

If we start with a sequential pack and its requisite peaked $P(\theta)$ and then change it to have a flat distribution, the light cones disappear; as in the isotropic packs, the force distribution becomes too irregular to characterize. However, if we start with an isotropic pack and make the angle distribution peaked, we do not create light cones. Likewise, if we take the sequential pack with the flattened distribution from before and then try to restore the original contact distribution, the light cones do not reappear. There is evidently some other aspect of the pack which is being destroyed by this procedure. This other aspect must be due to the isostatic nature of the pack, since in a hyperstatic pack with a similarly peaked angle distribution, the response is a single large peak directly under the forcing point, rather than light cones. It seems that both isostaticity and a peaked $P(\theta)$ are necessary, but insufficient, to cause light cones.

\section{\label{Sec:Conclusion}Conclusion}

In summary, we have shown that the set of jammed isostatic systems is richer than formerly thought. There appear to be at least three different types of force propagation possible: the previously observed light cones which appear in sequential packs; a more elastic-like response in stabilized isotropic systems; and a non-continuum behavior where no average stress can be found. The light cones are not visible in all isostatic packs because the null stress condition holds for only a subset of them. The previous observations of the null stress law seem to depend not only on the isostatic nature of the packs, but also on some further feature which appears related to anisotropy in the contact angle distribution.

Our simulations were motivated by the desire to probe the marginally jammed state in ways which are difficult for experiments. Many experiments which consider the response to point probes use macroscopic particles such as photoelastic disks or grains of sand. Such systems tend to be hyperstatic rather than isostatic, have frictional forces, and do not allow for tensile contacts \cite{Amar2011, Geng2003, Geng2001, Reydellet2001}. Other experiments look at crystalline, rather than disordered, packings \cite{Mueggenburg2002}. Still others are more interested in the resulting particle rearrangements \cite{Habdas2004} or the critical scaling close to isostaticity \cite{Katgert2010} than they are in the force response. 

The existence of friction in experimental systems is one barrier to performing real-world tests of our results. In the presence of these extra forces, there is no longer a contact number directly correlated with the onset of pack stability \cite{Somfai2007}. This may affect the force propagation at isostaticity. Frictional forces may also decrease the range of length scales over which hyperbolic behavior can be observed in packs which are only slightly hyperstatic. Their presence tends to make the response as a whole more elastic-like \cite{Goldenberg2005}. This barrier can be overcome by, for example, using a colloidal system where the beads do not directly touch, eliminating any contact friction \cite{Tan2012,Zhou2006}.

We allowed tensile contacts in our systems based on the discussion in Section~\ref{Sec:Background} which noted that the presence of light cones in sequential packs is not affected by restrictions on the sign of the contact forces. However, the sign may be important in the non-sequential packs. One reason the non-continuum behavior described in Section~\ref{Sec:IsotropicPacks} is possible is that arbitrarily large forces are allowed; we can have a large positive force since there can be a negative force of a similar size to balance it. If no tensile contacts are allowed, then individual contact forces are limited in size by the magnitude of the applied external force. Unfortunately, because there is a unique solution to Equation~\ref{Eqn:ForceBalance}, it is not possible to restrict the sign without making some fundamental changes to the packing geometry. Such changes did not affect the light cones in the sequential packs, but they may affect the results in the systems which did not average. Creating isostatic packs with tensile forces in a laboratory setting is also difficult. Tensile forces imply an attractive potential, which will tend to cause particles to clump together and produce more contacts than would be present in an isostatic pack. An additional experimental complication is that the arbitrarily large contact forces which are allowed can drive a rearrangement of the system. Indeed, even if there are no tensile contacts, the system will still be quite delicate, and small applied forces will tend to lead to small rearrangements. As the applied forces get larger, this can qualitatively change the distribution of contact forces \cite{Zhou2010}. 

Despite the difficulties in observing these effects in a laboratory setting, understanding the behavior of these systems is still important for the analysis of packs which are more readily observed. Even a system which is hyperstatic can display isostatic behavior on small enough length scales. Furthermore, any transition from a dilute system to a densely packed one must pass through the isostatic state.

Further work should be done to pinpoint what conditions lead to the existence of null stress behavior. Both an anisotropic contact angle distribution and isostaticity seem to be necessary but insufficient, as evidenced by the experiments in Section~\ref{Sec:ContactAngles} where the contact angles were changed while retaining the isostatic nature of the packs. We would like to improve our understanding of this issue by finding a better way to create non-lattice configurations with different contact angle distributions than the rather intrusive method described. It is possible that any sort of anisotropy in the pack construction will yield light cones. In this case, it would be interesting to to examine the connection between the light cones and network anisotropies to discover how the crossover from isotropic to anisotropic behavior occurs.

A surprising result is that boundary modifications can evidently change the bulk force propagation from non-continuum to elastic, even at length scales smaller than the sample size, and even though only a vanishingly small fraction of the vibrational modes are affected. There must be a way to understand the contrasting force responses in terms of these modes. We have begun preliminary investigations into the structure of the modes which are formed when an isostatic system is disconnected from its boundaries. These are a promising lead toward an explanation of both the presence of light cones and the dependence on the boundaries.

\appendix
\section{\label{Sec:DMMTProof}Proof that $\mathcal D = MM^T$}

In Section~\ref{Sec:Background}, we claimed that the dynamical matrix $\mathcal D$ and the contact matrix $M$ are related via $\mathcal D=MM^T$, and offered a proof that depended on the invertibility of $M$. However, this relationship is true even if $M$ is not invertible, as will be shown here.

In the following, Greek indices will be used to denote contacts. Other indices will be written as a pair giving first the bead number and next the vector component. For example, for the displacement vector $\nkv u$, $u_{ix}$ gives the x component of the displacement of the $i$th bead from its initial position. We also define $c_{ij}$ to be 1 if beads $i$ and $j$ are in contact, and 0 otherwise. $c(i)$ is a list of the beads in contact with $i$; that is, all $j$ such that $c_{ij}=1$. 

To begin, we will look at the structure of $MM^T$, by computing $(MM^T)_{ix,jx}$ when $i\ne j$. This is just
\[
(MM^T)_{ix,jx}=\sum_\alpha M_{ix,\alpha}M_{jx,\alpha}
\]
From Section~\ref{Sec:Background}, we know that $M_{ix,\alpha}=\cos\theta_{ik}$ if $\alpha$ is the contact between beads $i$ and $k$, and 0 otherwise. Thus $M_{ix,\alpha}M_{jx,\alpha}$ is only nonzero if $i$ and $j$ are in contact, and $\alpha$ is the contact between them. Each pair of beads can have at most one contact between them, so the sum collapses and we get
\[
(MM^T)_{ix,jx}=\cos\theta_{ij}\cos\theta_{ji}c_{ij}=-\cos^2\theta_{ij}c_{ij}
\]
If we were looking at the $y$ components instead, we would get $-\sin^2\theta_{ij}c_{ij}$, and if we were looking at a mixture of $x$ and $y$ components, it would be $-\sin\theta_{ij}\cos\theta_{ij}c_{ij}$. 

If $i=j$, then $\sum_\alpha M_{ix,\alpha}M_{ix,\alpha}$ will have a nonzero term for every contact that $i$ has. We get 
\[
(MM^T)_{ix,ix}=\sum_{k\in c(i)}\cos^2\theta_{ik}
\]
As before, if we were looking at $y$ components we would have a $\sin^2\theta_{ik}$ term instead, and mixed components would give $\sin\theta_{ik}\cos\theta_{ik}$.

Now we want to compare this to the form of $\mathcal D$. From Equation~\ref{Eqn:DynamicalMatrix}, we can expand to see
\begin{align*}
\mathcal E =\frac k2 \nkv u^T\mathcal D\nkv u = \frac k2\sum_{i,j}\big( &u_{ix}u_{jx}\mathcal D_{ix,jx}+u_{iy}u_{jy}\mathcal D_{iy,jy}\\ \notag &+u_{ix}u_{jy}\mathcal D_{ix,jy}+u_{iy}u_{jx}\mathcal D_{iy,jx}\big)
\end{align*}
We split this into diagonal and off-diagonal parts, and modify the off-diagonal portion so that every ($i$,$j$) pair is only written once:
\begin{align}
\label{Eqn:MultiplyDOut}
\mathcal E &= \frac k2\sum_{i<j}\big(2u_{ix}u_{jx}\mathcal D_{ix,jx}+2u_{iy}u_{jy}\mathcal D_{iy,jy}\\ \notag &+2u_{ix}u_{jy}\mathcal D_{ix,jy}+2u_{iy}u_{jx}\mathcal D_{iy,jx}\big)\\ \notag
&+\frac k2 \sum_i\big(u_{ix}^2\mathcal D_{ix,ix}+u_{iy}^2\mathcal D_{iy,iy}+2u_{ix}u_{iy}\mathcal D_{ix,iy}\big)
\end{align}

Here we have used the fact that $\mathcal D$ is symmetric, meaning $\mathcal D_{ix,jy}=\mathcal D_{jy,ix}$.

Next we compute the energy between a pair of beads $i$ and $j$ when they are displaced some small amount. If they were not in contact before this displacement, then there is no interaction energy. If they were, then the energy is given by 
\begin{equation}
\label{Eqn:HarmonicEnergy}
\mathcal E_{ij}=\frac k2\Big(r_{ij}-R_{ij}\Big)^2
\end{equation}
where $\nkv R_{ij}$ is the vector pointing from bead $i$ to $j$ before the displacements $\nkv u_i$ and $\nkv u_j$ have been taken into account, and $\nkv r_{ij}=\nkv R_{ij}+\nkv u_j-\nkv u_i$ is the same vector after the displacements have been made. The length of $\nkv r_{ij}$ is given by
\[
r_{ij}^2=\Big(R_{ij}\cos\theta_{ij}+u_{jx}-u_{ix}\Big)^2+\Big(R_{ij}\sin\theta_{ij}+u_{jy}-u_{iy}\Big)^2
\]
Taking the square root and assuming that the $u_i$ are small compared to $R_{ij}$, we see
\[
r_{ij}=R_{ij}+\cos\theta_{ij}(u_{jx}-u_{ix})+\sin\theta_{ij}(u_{jy}-u_{iy})
\]

Putting this into Equation~\ref{Eqn:HarmonicEnergy}, we can sum over all pairs of beads to get the total energy, and expand it into the form
\begin{widetext}
\begin{align}
\label{Eqn:ExpandEnergy}
\mathcal E &= \frac k2\sum_{i<j}\Big(2u_{ix}u_{jx}(-\cos^2\theta_{ij}c_{ij})+2u_{iy}u_{jy}(-\sin^2\theta_{ij}c_{ij})\\ \notag
&+2u_{ix}u_{jy}(-\sin\theta_{ij}\cos\theta_{ij}c_{ij})+2u_{iy}u_{jx}(-\sin\theta_{ij}\cos\theta_{ij}c_{ij})\Big)\\ \notag
&+\frac k2\sum_{i<j}\Big((u_{ix}^2+u_{jx}^2)\cos^2\theta_{ij}c_{ij}+(u_{iy}^2+u_{jy}^2)\sin^2\theta_{ij}c_{ij}\\ \notag
&+2(u_{ix}u_{iy}+u_{jx}u_{jy})\sin\theta_{ij}\cos\theta_{ij}c_{ij}\Big)
\end{align}
\end{widetext}

Now consider one term of the second sum,
\[
S_{xx}=\sum_{i<j}(u_{ix}^2+u_{jx}^2)\cos^2\theta_{ij}c_{ij}
\]
This is essentially a sum over contacts; for every $(i,j)$ pair in contact we get two terms with a factor of $\cos^2\theta_{ij}$: one for bead $i$, and one for bead $j$. We can write this instead as a sum over beads, using the fact that $\cos^2\theta_{ij}=\cos^2\theta_{ji}$:
\[
S_{xx}=\sum_i u_{ix}^2\sum_{k\in c(i)}\cos^2\theta_{ik}
\]
The other two terms in the final sum of Equation~\ref{Eqn:ExpandEnergy} can also be rewritten in this manner. Since this is true for all small vectors $\nkv u$, comparing the result with Equation~\ref{Eqn:MultiplyDOut} gives us the terms of the dynamical matrix. We find
\begin{align}
\mathcal D_{ix,ix}&=\sum_{k\in c(i)}\cos^2\theta_{ik}\\ \notag
\mathcal D_{iy,iy}&=\sum_{k\in c(i)}\sin^2\theta_{ik}\\ \notag
\mathcal D_{ix,iy}&=D_{iy,ix}=\sum_{k\in c(i)}\sin\theta_{ik}\cos\theta_{ik}\\ \notag
\mathcal D_{ix,jx}&=-\cos^2\theta_{ij}c_{ij}\\ \notag
\mathcal D_{iy,jy}&=-\sin^2\theta_{ij}c_{ij}\\ \notag
\mathcal D_{ix,jy}&=D_{iy,jx}=-\sin\theta_{ij}\cos\theta_{ij}c_{ij}
\end{align}
These are exactly the same as the the elements of $MM^T$ that we found earlier. Thus even if $M$ is not invertible, we still have $\mathcal D=MM^T$.

\section{\label{Sec:GoodFloors}Prevalence of boundaries which preserve isostaticity}

Here we give an argument to support the claim from the end of Section~\ref{Sec:IsotropicPacks} that even as the number of beads in a pack gets large, one can always expect to find a floor which does not create a contact matrix $M$ with a null space. To do this, we estimate the probability that a given floor will produce no null vectors, and compare this to the total number of floors possible. 

Looking back at Figure~\ref{Fig:ModifyFloor}, we can note that the nonzero elements of the null vectors are localized not only close to the floor, but also in different, possibly overlapping, horizontal areas. By deforming the floor in one place, we can eliminate the null vector located there and decrease the dimension of the null space without much affecting what happens along a different part of the floor. This indicates that it is some particular local configurations of floor beads which create the null vectors. We can model this by assuming that there is some probability per unit length $\alpha$ of a floor having a configuration which creates a null vector. 

Under this model, if many floors of a given length $\ell$ are examined, the dimensions of their null spaces should follow a Poisson distribution with mean equal to $\alpha \ell$. This is precisely what is shown in Figure~\ref{Fig:NumberNullVectors}. Using the fact that $\ell\sim\sqrt{N}$, we find that $\alpha \approx 0.3$ for all pack sizes. In units of bead-diameters, if the floor has length $\ell$, then the probability that there will be no null vectors is
\begin{equation}
\label{Eqn:ProbGoodFloor}
P(\textrm{no null vectors})=\left(\begin{array}{c}\ell\\0\end{array}\right)\alpha^0(1-\alpha)^\ell=(1-\alpha)^\ell
\end{equation}

We next estimate the number of floors possible in a given pack. We limit ourselves to floors consisting of a chain of beads in contact with each other which wraps around the horizontal periodic boundary exactly once before looping back on itself.  Given some starting bead, we can approximate this as a one dimensional random walk where each step moves one bead to the right, and then randomly up or down. This is a crude approximation of the geometry in an isostatic pack, where each bead has on average four contacts; on either side, it will tend to have one contact higher in the pack, and one lower. This is exactly the situation for the diagonal isostatic lattice shown in Figure~\ref{Fig:IsostaticLattices}a. 

After $\ell$ such steps, our path must return to the horizontal position of the starting bead, within one vertical step of the original height. In a one dimensional random walk, the number of paths of length $\ell$ which end at height  $k$ is
\[
N_\ell(k) = \left(\begin{array}{c}\ell\\ (\ell+k)/2\end{array}\right)
\]
Using Stirling's formula, we find that the total number of possible floors $N_F$ is
\begin{align*}
N_F&=\left(\begin{array}{c}\ell\\ (\ell+1)/2\end{array}\right)+\left(\begin{array}{c}\ell\\ \ell/2\end{array}\right)+\left(\begin{array}{c}\ell\\ (\ell-1)/2\end{array}\right)\\ &\approx 3\sqrt{\frac{2}{\pi \ell}}2^\ell
\end{align*}
Combining this with Equation~\ref{Eqn:ProbGoodFloor}, we can compute the expected number of floors with no null space $E_0$ to be 
\[
E_0 \sim (1-\alpha)^\ell \ell^{-1/2} 2^\ell
\]
Since $\alpha<1/2$, this grows with increasing path length. The path length $\ell$ must grow at least as fast as the linear size of the system, so we can take $\ell\sim\sqrt{N}$. Thus the expected number of floors that do not create a null space in $M$ grows with increasing system size.

\section{\label{Sec:ElasticTheory}Stress fields in elastic media}

Here we state the results of Leonforte, et. al. \cite{Leonforte2004} which we used in Section~\ref{Sec:HardTop} to compare the stresses in isotropic packs with added hard ceilings with the stresses in an elastic solid. 

We will use an $L\times L$ box centered at the origin which is periodic in the $x$ direction. The method is to solve the equations for linear elasticity assuming perfectly rigid hard ceilings and bottoms that are very rough, so that the displacements $\nkv u(x,\pm L/2)$ there are zero. The external force is modeled as a vertical external pressure $p(x)$ in the middle of the pack. The medium is divided into two parts: part 1, for $y>0$ and part 2, for $y<0$. The additional boundary conditions are then $\nkv u^{(1)}(x,0)=\nkv u^{(2)}(x,0)$, $\sigma_{yy}^{(1)}(x,0)=\sigma_{yy}^{(2)}-p(x)$, and $\sigma_{xy}^{(1)}(x,0)=\sigma_{xy}^{(2)}(x,0)$. The finite boundaries quantize frequencies to multiples of $\Delta q = (2\pi/L)$, and we can write the pressure as 
\begin{equation}
p(x) = \sum_{n=0}^\infty \cos(q_nx)s(q_n)\Delta q
\end{equation}
where $q_n = n\Delta q$.

Instead of using a point force, we will choose $p$ to be a Gaussian that is narrow enough that almost the entire area is contained in a length of approximately the same size as a bead diameter. Modest changes to this width do not much affect the results.

Solving for the stress gives
\begin{widetext}
\begin{align*}
\label{Eqn:TheoryStress}
\sigma_{xx}^{(1,2)}+\sigma_{yy}^{(1,2)}&=\sum_{n=0}^\infty\cos(q_n x)\big[a^{(1,2)}e^{q_n y}+b^{(1,2)}e^{-q_n y}\big]\\ \notag
\sigma_{xx}^{(1,2)}-\sigma_{yy}^{(1,2)}&=\sum_{n=0}^\infty\cos(q_nx)\Big(q_ny\big[a^{(1,2)}e^{q_ny}-b^{(1,2)}e^{-q_ny}\big]+2\big[c^{(1,2)}e^{q_ny}-d^{(1,2)}e^{-q_ny}\big]\Big)\\ \notag
\sigma_{xy}^{(1,2)}&=\sum_{n=0}^\infty\sin(q_nx)\Big(\frac{q_ny}{2}\big[a^{(1,2)}e^{q_ny}+b^{(1,2)}e^{-q_ny}\big]+\big[c^{(1,2)}e^{q_ny}+d^{(1,2)}e^{-q_ny}\big]\Big) \\ \notag
a^{1}(q)&=-\frac{s(q)\Delta q}{4\delta (q)}\Big(\big[(1+\nu)g L + 3-\nu\big]e^{-qL}+(\nu-3)e^{-2qL}\Big)\\  \notag
a^{2}(q)&=\frac{s(q)\Delta q}{4\delta(q)}\Big(\big[(1+\nu)qL-3+\nu\big]e^{-qL}-(\nu-3)\Big)\\ \notag
b^{1}(q)&=\frac{s(q)\Delta q}{4\delta(q)}\Big(\big[-(1+\nu)qL+3-\nu\big]e^{-qL}+(\nu-3)\Big)\\ \notag
b^2(q)&=\frac{s(q)\Delta q}{4\delta(q)}\Big(\big[(1+\nu)qL+3-\nu\big]+(\nu-3)e^{-2qL}\Big)\\ \notag
c^1(q)&=\frac{s(q)\Delta q}{8(1+\nu)\delta(q)}\Big(\big[(1+\nu)^2q^2L^2/2+(1-\nu^2)qL+(1-\nu)(3-\nu)\big]e^{-qL}\\ \notag
&+(1-\nu)(\nu-3)e^{-2qL}\Big) = d^2(q)\\ \notag
c^2(q)&=\frac{s(q)\Delta q}{8(1+\nu)\delta(q)}\Big(\big[(1+\nu)^2q^2L^2/2-(1-\nu^2)qL+(1-\nu)(3-\nu)\big]e^{-qL}\\ \notag
&+(1-\nu)(\nu-3)\Big) = d^1(q))\\  \end{align*} \begin{align*} \notag
\delta (q)&=2Le^{-qL}-\frac{\nu-3}{2(\nu+1)}\Big(1+e^{-2qL}\Big)
\end{align*}
\end{widetext}

\begin{acknowledgements}
The author would like to thank Martin van Hecke for useful discussions, along with Martin Lenz and Efraim Efrati. This work is based on a PhD thesis at the University of Chicago under the supervision of Thomas A. Witten, and was supported in part by the National Science Foundation's MRSEC Program under Award Number DMR 0820054.
\end{acknowledgements}

\bibliography{../Bibliography/GranularBibliography.bib}

\end{document}